\documentclass[12pt,letterpaper]{article}

\usepackage[a4paper,total={7in,10in}]{geometry}

\usepackage{graphicx}
\usepackage{helvet}
\usepackage{authblk}
\usepackage{hyperref}
\usepackage{amsmath}
\usepackage{amssymb}
\usepackage{booktabs}
\usepackage{orcidlink}
\usepackage[super,comma,sort&compress]{natbib}
\graphicspath{{./}}

\makeatletter
\renewcommand{\maketitle}{\bgroup\setlength{\parindent}{0pt}
\begin{flushleft}
  \textbf{\@title}

  \@author
\end{flushleft}\egroup}
\makeatother

\title{Typical-query energy figures mislead institutional AI electricity demand}
\date{}

\author[1,2,\orcidlink{0000-0003-0343-0199},*]{Nathan A. Garland}
\author[1,3,\orcidlink{0000-0001-7626-1248}]{Ryan F. Heneghan}

\affil[1]{School of Environment and Science, Griffith University, Nathan 4111, QLD, Australia}
\affil[2]{Queensland Quantum and Advanced Technologies Research Institute, Griffith University, Nathan 4111, QLD, Australia}
\affil[3]{Australian Rivers Institute, Griffith University, Nathan 4111, QLD, Australia}
\affil[*]{Correspondence: n.garland@griffith.edu.au}

\begin{document}

\maketitle


\section*{HIGHLIGHTS}

\begin{itemize}
\item Open bottom-up model links LLM token geometry to institutional electricity
\item Highest-energy 10\% of requests draw 60\% of modelled inference electricity
\item Typical Wh $\times$ visible prompts misses 70-93\% of modelled annual electricity
\end{itemize}


\section*{CONTEXT \& SCALE}

The electricity footprint of AI inference using frontier large language models is often discussed through a single ``typical'' energy-per-query value. However, large institutions, such as government departments or universities, deploy AI as heterogeneous collections of chat, reasoning, retrieval, coding, and increasingly agentic workflows. Here, we develop a model framework to capture how this individual-level variability and uncertainty scales to institution-scale energy usage and  associated carbon emissions attributed to the use of frontier class AI, using Australian public universities as a case study. The highest-energy 10\% of individual-level requests can contribute approximately 60\% of institution-level energy footprint, and in a scenario where AI uptake continues modest acceleration to 2030, total model calls could be about twice the visible user-submitted messages. Critically, we find that scaling up a single median Wh-per-query value to estimate an institution's total electricity consumption from AI use can be wrong by between a factor of two and six, highlighting the need to move away from single energy-per-query assessments.

While present day institutional energy attributions due to LLM inference are not a large fraction of national power systems, the significance is instead the combination of rapid potential scaling, a potential shift in individual-level use to computationally intensive reasoning and agentic workflows, poor visibility of AI energy usage to institutional customers, and large variation in carbon consequence with electricity generation methods. Our framework therefore motivates task-specific provider disclosure and energy-aware accounting and reporting in AI as a service procurement by institutions.

\newpage 

\section*{SUMMARY}

Institutions increasingly buy large language model services, yet we still often estimate electricity use as single Wh-per-query value. We present the Surrogate Large Language Energy Emulator (SULLEE), an open bottom-up model linking token geometry, serving hardware, and workflow structure to annual institutional electricity, with uncertainty propagated by Monte Carlo simulation. When an institution multiplies a single reported "typical" per-query energy figure by its count of visible user prompts our modelling implies that this compounds two separable shortfalls: a right-skewed task-energy distribution not represented by any single scalar, and workflow amplification of each visible prompt into additional agent and tool calls. Together these imply an underestimate of total inference electricity by a factor of approximately 3.4 under present-day observed use, rising to approximately 15 under a 2030 high-integration scenario.  Using Australian public universities as a case study, we show how the top 10\% of highest-energy requests contributes over half of modelled institution-level energy use. We quantify how institution-level electricity demand may increase across three AI integration scenarios of: 1) a continuation of estimated present-day 2026 usage; 2) a high-uptake present-day 2026 institution, and 3) a potential 2030 uptake scenario - based on present trends of generative AI tools and their uptake. For a representative Australian university, mean annual electricity demand for AI inference purposes is projected to increase from 17.3~MWh (with a 3.1--47.1~MWh uncertainty, within P05--P95 scenario quartile ranges) under the 2026 observed-use scenario to up to 287.4~MWh (41.5--830.7~MWh) by 2030. 





\section*{KEYWORDS}
large language models, AI, energy, emissions, higher education


\newpage
\section*{INTRODUCTION}

\label{sec:introduction}

The deployment, and adoption, of large language models (LLMs) at scale over the past few years by institutional and corporate organisations raises significant questions regarding energy consumption and environmental impact~\citep{patterson2021,luccioni2023}.

The need for transparent, bottom-up (i.e., how individual-level AI use drives total institutional AI energy demand) modelling arises from limited availability of verified operational data from AI service providers. Major AI providers have historically not disclosed per-query energy metrics, making independent assessment difficult. Only recently have disclosures emerged, such as OpenAI's CEO quoting on a blog post 0.34~Wh per average ChatGPT query~\citep{altman2025}, while Google reported 0.24~Wh for median Gemini text prompts~\citep{elsworth2025}, and Microsoft Research reported a 0.31~Wh median~\citep{oviedo2025}. These isolated data points, while valuable, differ in statistical definition, workload boundary, and methodological transparency, and represent specific model and serving configurations that may not generalise or otherwise provide sufficient guidance for institutions wishing to understand and benchmark the aggregate impact of their LLM adoption. 

The distinction has practical consequences for how institutions interpret these figures. In a recent Washington Post piece~\cite{DunnZhao2026}, writing about student use of chatbots in a sustainability course, Dunn and Zhao compare the electricity used by a \textit{median} query with one second of microwave operation, then observe that ``Asking a bot 100 questions is roughly the equivalent of heating up a burrito''. However, estimating the electricity associated with 100 questions requires knowing how many model calls they generate and the \textit{mean} energy of those calls across the variety of tasks performed.

This challenge has been discussed in detail in prior work, notably by Jegham~\textit{et al.}~\citep{jegham2025}. Complementary simulation-based approaches have also emerged, for example \citet{ozcan2025} extend a high-fidelity LLM inference simulator with a graphics processing unit (GPU) power model and integrated grid-aware carbon-intensity co-simulation. However, such frameworks are designed to evaluate specific deployment hardware configurations and not to support generalised, technology-agnostic institutional planning. Key challenges to informing institutional-level AI use include:
\begin{enumerate}
\item \textbf{What is the ``average'' query?} Varying and often vague query definitions (a single exchange vs. full conversation).
\item \textbf{Proprietary architectures} Generative AI model architectures and serving configurations of proprietary models provided by frontier-class companies remain undisclosed for commercial interests, reducing transparency and community understanding of compute energy requirements.. 
\item \textbf{Variable deployment conditions} Energy consumption varies with data centre location, cooling systems, hardware generations, batch sizes, and utilisation patterns which are factors rarely disclosed.
\end{enumerate}

Recent analysis has also highlighted that test-time compute scaling, where models expend additional inference floating point operations on chain-of-thought reasoning, introduces a qualitatively new dimension to inference energy growth that is not captured by static model-size scaling alone~\cite{oviedo2025}. Despite inherent uncertainties of input parameters leading to wide uncertainties in potential outputs, coarse estimates remain valuable for institutional planning~\citep{strubell2019} - especially when otherwise more accurate estimates are not transparently available by providers. Order-of-magnitude estimates can help to inform budgeting for sustainability reporting, enable scenario analysis for policy development, or provide baselines against which future disclosures can be compared. Together, these documented gaps motivate a bottom-up, usage-based approach. 

Here, we present {{SULLEE}}~(\textbf{Su}rrogate \textbf{L}arge \textbf{L}anguage \textbf{E}nergy \textbf{E}mulator), a transparent mechanistic inference model of institution-level AI energy demand. SULLEE resolves institution-specific activity drivers (user counts, adoption scenarios, and workload and workflow intensity) to transparent assumptions, and makes it possible to compare outputs against emerging prompt-level measurement disclosures (e.g., Google’s published methodology and median prompt energy/emissions estimates for Gemini~\cite{elsworth2025}). In contrast with median estimates of prompt energy use, SULLEE draws from a distribution of individual-level prompt energy demand. At the same time, industry disclosures and scaling trends allow for sufficiently accurate representative surrogates of contemporary frontier model architecture scenarios, not exactly assumed technology of any proprietary commercial model. Finally, SULLEE also captures the impact of variable deployment conditions on institution-level energy demand. We demonstrate SULLEE's capability to produce defensible bounds on institution-level AI energy demand, which can enable sensitivity analyses, and forward projections that are immediately useful for governance and procurement - and that also serve as an external benchmark against which future vendor disclosures can be compared and incorporated. In this work we frame our application to institutional scale-up of inference task energy through a case study of Australian universities. We use a representative 20,000 student Australian university as the most frequent unit for discussion and analysis of energy and emissions, though where useful and relevant we also present quantities for a larger 40,000 student representative Australian university, as well as the total Australian university sector combined.

The contribution of this work is not to claim a new universal value for ``energy per query'', but a modelling hierarchy connecting three scales that are usually treated separately: (i) request-level inference computing mechanisms and serving constraints; (ii) heterogeneous human- and machine-generated workflow demand; and (iii) institution-scale electricity and carbon accounting. Relative to production measurements and deployment-level inference analyses \cite{elsworth2025,oviedo2025}, SULLEE provides a transparent surrogate for otherwise unobservable frontier-serving configurations, while the institutional layer makes dependence on adoption, task mix, and workflow amplification explicit. This separation also allows measured quantities, modelling approximations, epistemic priors, and deterministic sensitivities to remain distinguishable instead of a single per-query coefficient.

\section*{METHODS: SULLEE inference model}\label{sec:methods}

\subsection*{Study design and modelling hierarchy} \label{sec:methods-study-design}

\begin{figure}
    \centering
    \includegraphics[width=0.9\linewidth]{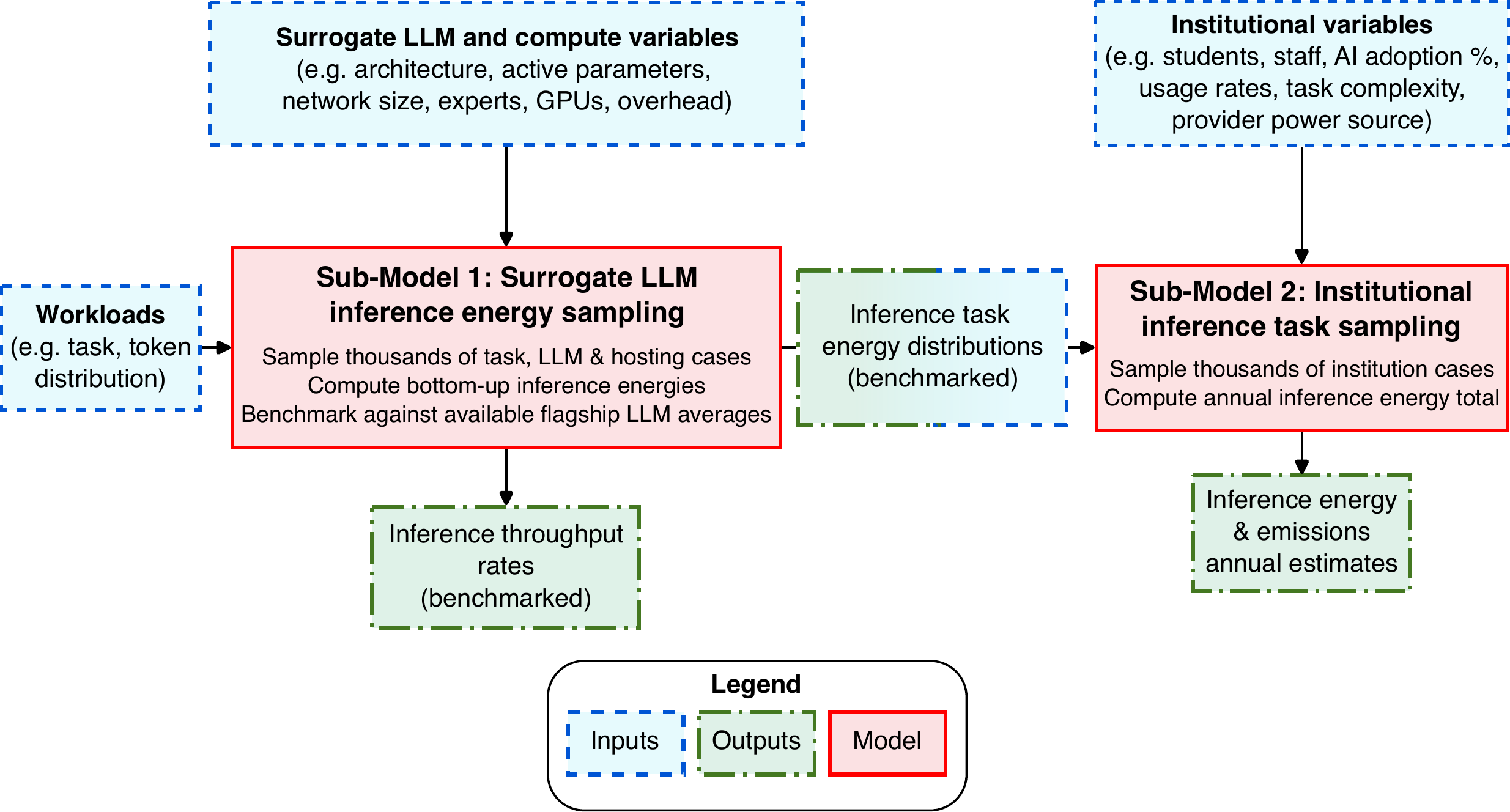}
    \caption{The conceptual framework of this modelling study, showing the interconnection between a base-level inference energy model and the higher-level scale up simulation of institutional scenarios.}
    \label{fig:figure1}
\end{figure}

SULLEE is comprised of two sub-models, summarised in Figure \ref{fig:figure1}. Sub-model 1, the Surrogate LLM inference energy sampling sub-model, estimates the energy required to process a single LLM inference request with representative frontier class transformer based architecture. It combines a individual-level workload requests with the characteristics of a representative hyperscaler serving site. The approach is intentionally bottom-up to avoid relying on a specific provider-reported aggregate energy number. We approximate the dominant physical and architectural contributors to transformer inference, such as active parameter count, prompt length, output length, attention-related memory traffic, hardware throughput, and site-level overhead. Output from this first sub-model is then amplified at institutional scale by the second sub-model, using a Monte Carlo simulation to estimate annual inference electricity footprints, obtained by summing the modelled LLM-serving energy of all included human-initiated and machine-initiated workflow streams. 


\section*{Sub-model 1: Surrogate LLM inference energy sampling} \label{sec:methods-energy-boundary}

The fundamental inference request unit in SULLEE is one completed LLM text-generation request with explicitly specified input and output token counts, $(T_{\rm in},T_{\rm out})$. The model's task is to then calculate the primary request-scale estimand of required operational energy (in Wh) for a single completed request. In this study, we focus on modelling a surrogate of closed source frontier-class LLM tools used by consumers and enterprises alike.

We estimate operational energy attributable to LLM \emph{serving} (inference) through allocated IT-systems like accelerator/host service power, $P_{\rm IT}$, multiplied by a modelled active service time of the inference task. We do \emph{not} model training energy, embodied hardware emissions, end-user device power, or upstream telecommunications outside the declared boundary.

Given a $P_{\rm IT}$ requirement of the GPU/accelerator/host unit handling an inference task with explicitly specified input and output token counts, $(T_{\rm in},T_{\rm out})$, it may be scaled by the time taken to process both input and output tokens to obtain an energy value.  For each request the physical runtime is decomposed into prefill and decode phases of inference, with throughput processing time per token, $t_{\rm pre}$ and $t_{\rm dec}$, determined by a token-time model based on the transformer architecture and hardware serving the model. These terms include parameter-associated compute, architecture-specific non-parameter compute, memory traffic, and internal communication. Batching, deployment topology, and resource allocation are explicit in the token throughput serving calculation with details provided in the subsequent sections as well as the supplementary material. The active IT energy,  $E_{\rm active,IT}$, can be represented as

\begin{equation}\label{eq:main_energy}
E_{\rm active,IT}
=
\frac{P_{\rm IT}}{3600}
\left(
T_{\rm in}t_{\rm pre}
+
T_{\rm out}t_{\rm dec}
\right),
\end{equation}
where $P_{\rm IT}$ is measured in watts, token times in seconds, and dividing by a factor of 3600 allows the energy  $E_{\rm active,IT}$ to be presented in units of watt-hours.

\paragraph{Power overhead and hardware parameters} This direct $E_{\rm active,IT}$ quantity is scaled by a classical data-centre overhead Power Usage Effectiveness (PUE), and a separate aggregate fleet-accounting factor~\cite{oviedo2025}.  A production fleet factor $\beta_{\rm{fleet}} $ represents deployment-scale non-uniform demand, networking and redundancy that are not represented by active request service alone~\cite{oviedo2025}; it is not an intrinsic GPU-power or throughput multiplier. The assumed delivered fleet boundary inference energy for a single completed request is then

\begin{equation}
E_{\rm fleet}
=
E_{\rm active,IT}\,
{\rm PUE}_{\rm prod}\,
\beta_{\rm fleet},
\qquad
\beta_{\rm fleet}=1.31.
\end{equation}

Following Oviedo et al.~\citep{oviedo2025}, we adopt a reference fleet factor of 1.31, comprising approximate allowances for non-uniform daily demand (1.08), network-fabric electricity (1.10), and redundant provisioned capacity (1.10). The networking allowance is informed by NVIDIA DGX SuperPOD reference-system power specifications, while the demand and redundancy terms represent illustrative fleet-operating assumptions rather than directly measured properties of a specific commercial provider. The factor is applied once and accounts only for electricity not already represented within the active-node calculation or PUE.

For active-node power attribution, we adopt a steady-state reference of 70\% of the manufacturer's specified maximum node power, with 40\% and 90\% retained as deterministic sensitivity cases. This choice follows the modelling convention of Oviedo et al.~\citep{oviedo2025} and is supported qualitatively by empirical studies demonstrating that LLM inference power varies substantially with request phase, batch configuration and workload, and generally remains below rated peak power~\cite{mlenergy2025,Patel2024}. It is not interpreted as a measured mean for NVIDIA H200 deployments.

Production PUE is represented by a physically constrained epistemic log-normal prior with nominal fifth and ninety-fifth percentiles of 1.05 and 1.40, respectively. This range follows Oviedo et al.~\citep{oviedo2025} and encompasses efficient hyperscale operating conditions reported by major cloud providers, while permitting higher facility overhead. The distribution describes plausible hosting configurations and not an empirically established population distribution of provider PUE values. Together, these assumptions define a transparent fleet-attribution scenario.

\subsection*{Inference workloads and task-conditioned token use}
\label{sec:methods-workload}

Besides LLM architecture and hosting configuration, an important input of this modelling process is the length or size of inference tasks being performed. Any LLM inference request submitted by a user is interpreted and fragmented into \textit{tokens}. Tokens often correspond to common word fragments (e.g., “un-”, “break”, “able”), or whole words, punctuation, whitespace, or more. Prior research has noted that prompt token lengths are characterised by a large majority of inference task lengths bunched around a median value that is relatively modest, accompanied by a long tail of tasks of much longer token lengths that indicate the ability of user requests to grow very long. These large token length tasks are known to occur with a decreasing frequency as these very long requests are not as common~\cite{wang2025burstgpt,she2026laps}. Rather than having to pick specific inference task sizes, we can draw from a probability distribution of inference tasks that users may ask of an LLM to better approximate the reality of a continuum of task sizes.

Modelling this variation through the use of broad log-normal input distributions is further motivated by empirical patterns that have been long observed in human-generated text, content, and data in general~\cite{Gros2012}. A prior study has suggested a Zipf-like distribution~\cite{wang2025burstgpt}, while Pareto distributions are often employed in long-tail statistics applications as well. Given the shared qualitative similarities between these long-tailed distributions, we employ task-conditioned log-normal input distributions as transparent first-order approximations instead of imposing a single provider-independent prompt-length distribution.

Consistent with trends in recent work~\cite{mlenergy2025}, we represent six request classes each with variable energetic loads, rather than using one generic request: \textit{general chat}, \textit{long document}, \textit{code completion}, \textit{Grounded retrieval augmented generation (RAG)}, \textit{reasoning production}, and \textit{deep problem solving}. The classes are intended to represent materially different token geometries; ordinary general, long-document, code, and RAG inputs are represented by positive log-normal distributions, while their outputs are represented by exponential distributions. Reasoning-production and deep-problem-solving inputs and outputs are represented by broader log-normal distributions to retain their observed long output tails \citep{servegen2026,wang2025burstgpt}.

Within a task class, input and output tokens, $T_{\rm in}$ and $T_{\rm out}$ respectively, are sampled conditionally independently. This is a first-order modelling approximation, not a claim that individual request and response lengths are physically independent. Dependence between request geometry and workload type is represented explicitly through the request class. To keep generated request sizes sensible, samples violating reasonable task-specific token limits, or a 128K token total context maximum are rejected and resampled. We choose 128K tokens as a cut-off maximum in line with the practical context windows offered by a range of contemporary frontier-class generative AI models. No arrival process, session process, or queuing model is introduced by this workload generator. 

The aggregate SULLEE request distribution, combining all six request classes, is a reference scenario distribution used to probe uncertainty across different user types as well as the sampled various model scenarios and not a single empirical estimate of the global distribution of commercial LLM traffic. The complete task-conditioned input/output distributions, truncation limits, evidence status, and reference-mixture assumptions are reported in Tables~S1A--S1B and Supplementary Equations~S1. The task-composition weights used to construct the request-scale reference population are illustrative and are not interpreted as measured market prevalence. Task-conditioned energy distributions are therefore the more portable model outputs. Quantities such as the pooled mean, median, tail shares, and the empirical percentile occupied by an external provider disclosure are conditional on the stated SULLEE reference mixture.

\subsection*{Closed-frontier inference model scenarios and architecture uncertainty}\label{sec:methods-architecture}

Rather than rely on specific open-source models, when building SULLEE we draw from disclosed LLM architectures as a basis and then project, or scale up, based on observed industry trends in scaling of frontier model sizes to propose an approximate, general \textit{surrogate} family of LLM architecture and hosting configurations. The surrogate sub-model within SULLEE is designed to emulate the general architecture characteristics of present frontier LLMs developed and hosted through hyperscaler providers purely for the purposes of benchmarking ranges of inference energy requirements across plausible LLM-usage scenarios.

We represent uncertainty in frontier-class language model architecture using four structural scenarios informed by publicly documented open-weight systems: Mistral Large 3 \citep{mistral2025large3}, Kimi K2.5 \citep{Kimi25}, DeepSeek V4 Pro \citep{DeepseekV4}, and Kimi K3 \citep{KimiK3}. These models span substantial variation in total and activated parameter counts, expert routing, network depth, and attention architecture. Their published specifications provide reference points for constructing four internally consistent surrogate inference configuration scenarios. Note, the scenarios are not intended to reproduce any individual provider's deployment or serving implementation, nor to identify the undisclosed architectures of proprietary frontier models. 

These scenarios provide physically consistent combinations of total and active parameter counts, expert topology, layer organization, attention or recurrent-state geometry, and deployment requirements. They embody plausible, representative LLM architecture structural modelling scenarios rather than claims of exact architectures of specific provider IP.

For the primary request-scale experiment, $N$ sampled workload request geometries are evaluated under each of the four structural transformer architecture scenarios. The resulting primary population therefore contains $4N$ request--architecture evaluations. Each structural scenario is assigned equal epistemic weight, $w_j=0.25$.

Architecture uncertainty represents persistent uncertainty about the serving model not independent request-to-request architecture switching. Therefore at institutional scale, architecture uncertainty is moved to the outer Monte Carlo loop: one of the four structural scenarios is sampled once per institutional realization and retained across the roles, tasks, and workflow streams in that realization. This is done to simulate a scenario where an institution may choose a particular frontier LLM provider to license and deploy to users. 

\paragraph{Interpretation of model input priors and sensitivities}
In this work the model inputs are typically either externally measured or disclosed quantities (e.g. a maximum power rating or RAM capacity of a GPU accelerator), or epistemic scenario priors used to represent quantities that are not generally publicly observable or reported (e.g. the number of parameters in a LLM, or the overhead PUE of a data centre). Monte Carlo sampling of an epistemic prior is a common method for propagating uncertainty within declared priors but does not imply that the sampled distribution is an empirically measured population distribution. Likewise, in our approximations of representative frontier-class models the equal weighting of the four closed-frontier structural scenarios is an epistemic coverage prior, and not an assumption of these models being the literal configurations employed by any particular frontier providers. As a result when analysing and interpreting output, we show reported P05--P95 ranges as quantiles for clear uncertainty of the specified scenario model unless explicitly stated otherwise.

\subsection*{Hardware deployment topology and assumptions}
\label{sec:methods-deployment}

In the estimation of inference energy in equation~\eqref{eq:main_energy}, for any given query we must know the power being drawn by the compute hardware doing the inference task. To determine this power consumption of the hardware, a dominant factor is understanding how many GPU accelerators, each with the same available maximum RAM, are required to store a large neural network's set of parameters or weights. In reality, AI inference also requires additional memory for token generation cache, running multiple user queries concurrently, and other runtime overhead, so the practical requirement is larger than the memory needed to store the model weights alone. Recent work has shown that multiple-user batching and long-context inference up to $10^5$ tokens can increase required GPU memory by factors of 2 or 3 beyond that needed for the model weights alone~\cite{davies2025}. The current SULLEE implementation therefore checks that the memory required for model weights, key-value (KV) cache, runtime workspace, and communication buffers can be accommodated on each GPU, allowing for uneven memory requirements across GPUs. It then selects the smallest feasible deployment in increments of eight GPUs per compute node.

When determining the memory required to store a large neural network's parameters, for the primary analysis we assume FP8 storage of model weights (i.e. 1 byte per parameter) as a representative default, while FP4 storage and BF16 are evaluated as sensitivity cases. Recent advances towards 4-bit quantisation without major inference quality degradation remain an advancing prospect for hyperscaler inference compute~\cite{shen2025,lin2024}.

The primary hardware reference is an eight-GPU NVIDIA H200 SXM node, with 141~GB high bandwidth memory (HBM) and 4.8~TB\,s$^{-1}$ HBM bandwidth per GPU. Hardware constants are taken from manufacturer specifications and stored in the model implementation and input files; these values are representative hardware assumptions rather than claims about the hardware used by a particular commercial provider.

For a candidate number of GPUs in a deployment (e.g. 8, 16, 32, and so on), the model checks the memory requirements of the most heavily loaded GPU in lieu of relying on the total available cluster memory. In the central memory scenario, the estimated model-weight memory per GPU includes a 5\% allowance for additional weight storage and accounts for uneven distribution of model layers across GPUs. Each GPU additionally reserves 2.0~GiB for runtime workspace and 0.5~GiB for communication buffers. Of the remaining HBM, 90\% is made available for KV cache and 10\% is retained as a runtime memory reserve.

For the four primary FP8 surrogate frontier-model scenarios, the resulting minimum feasible H200 deployments are 8, 8, 16, and 32 GPUs, respectively. Hardware constants, detailed GPU memory allocations, precision cases, and deployment rules are reported in Supplementary Equation~S5 and Tables~S2A--S2C.

\subsection*{Approximating serving time from throughput}
\label{sec:methods-serving}

As established in equation~\eqref{eq:main_energy}, given a set of tokens for the prefill/input, $T_{\rm in}$, and decode/output, $T_{\rm out}$, phases of an inference request for a model with $p_{\rm active}$ network parameters, we need to estimate the token-time (i.e. serving time per input and output token) of each phase, $t_{\rm pre}$ and $t_{\rm dec}$, based on the LLM architecture and hardware it is being served with. A schematic of the logical flow of information is shown below.
\[
\boxed{
\underbrace{p_{\rm active}}_{\text{model size}}
\;\longrightarrow\;
\underbrace{F_x,\ M_x}_{\text{arithmetic/memory operation requirements}}
\;\longrightarrow\;
\underbrace{R_x}_{\text{token throughput rates}}
\;\longrightarrow\;
\underbrace{t_x}_{\text{serving times}}
\;\longrightarrow\;
\underbrace{E}_{\text{energy}}
}
\]

These serving times are determined by the  reciprocal of the aggregate throughput
\[
t_x=\frac{1}{R_x},\qquad
x\in\{\mathrm{pre},\mathrm{dec}\},
\]
where \(R_x\) is the final aggregate throughput and \(t_x\) is the corresponding amortised serving time per token. To access these quantities, we estimate the throughput limits for prefill, $R_{\rm pre}$, and decode , $R_{\rm dec}$, with roofline-style~\citep{williams2009} performance bounds as demonstrated in prior LLM characterisation studies~\citep{yuan2024,zhang2023}. SULLEE modifies these bounds with explicit cache, batching, topology-aware communication, deployment, and power-allocation considerations. 

\paragraph{Roofline definition of throughput rates}
The estimated computational work needed for prefill and decode stages is converted to token throughput using compute and memory roofline limits. For either inference phase $x\in\{\mathrm{pre},\mathrm{dec}\}$, the corresponding throughput for a batch of $B$ requests is bounded by

\begin{equation}
R_x^{\rm roof}(B)
=
\min\left(
\frac{\eta_{\rm compute}\mathcal{C}_{\rm peak}}
     {F_x(B)},
\frac{\eta_{\rm HBM}\mathcal{B}_{\rm peak}}
     {M_x(B)}
\right),
\end{equation}
where $F_x$ and $M_x$ are the modelled floating point arithmetic operations and memory traffic per token, respectively, $\eta_{\rm compute}$ and $\eta_{\rm HBM}$ are efficiency factors of compute and memory processing in realistic serving hardware scenarios, and $\mathcal{C}_{\rm peak}$ and $\mathcal{B}_{\rm peak}$ are the corresponding peak hardware compute rate and memory bandwidth. The resulting serving throughput also accounts for exposed communication time and deployment constraints, as detailed in Supplementary Equation~S6.

In a prefill stage, the number of input tokens determines the compute load $F_\text{pre}$, while in decode to determine the number of additional floating-point operations required in an inference task, as a model's parameter count approximately equals the sum of weight matrix elements, the dominant parameter-associated contribution to the amount of compute operations needed may be written
\begin{equation}
    F_\text{dec} \approx 2 p_\text{active},
    \label{eq:flops_basic}
\end{equation}
where $p_\text{active}$ denotes parameters activated per token. For dense models $p_\text{active} = p_\text{total}$, while for sparse mixture-of-experts architectures $p_\text{active} < p_\text{total}$. 
\paragraph{Compute efficiency}
In this work, we decouple the efficiency of compute and memory operations based on recommended practices observed in literature~\cite{recasens2025}. A compute-roofline efficiency, $\eta_{\rm compute}$, is fixed at 0.40 as a representative effective fraction of theoretical peak arithmetic throughput. This assumption is informed by measured LLM inference performance, such as  prefill-attention utilisation of approximately 34–48\% reported by Ye et al. \citep{ye2025}, and the 43\% model floating point operations (FLOPs) utilisation reported by Pope et al.~\citep{pope2023} for a low-latency large-transformer prefill configuration. Higher efficiencies are achievable under optimised workloads and serving configurations \citep{Shah2024}. Accordingly, $\eta_{\rm compute}=0.40$ is treated as a literature-informed modelling assumption rather than a calibrated efficiency for a particular frontier model or H200 deployment.

\paragraph{HBM-efficiency calibration}
Consistent with literature observations~\citep{recasens2025} of modern large-scale transformer models often being memory-limited, we employ a high bandwidth memory (HBM) roofline efficiency of $\eta_{\rm HBM}=0.43$ determined by a calibration protocol against an empirical data set of recent open source models described below; no additional post-hoc throughput or energy fitting against provider measurements is applied. In order to quantify this uncertainty, and estimate an informed baseline value of $\eta_{\rm HBM}$ for our modelling, the bottom-up serving model is compared with 19 public NVIDIA TensorRT-LLM Hopper cases for which model identity, hardware, GPU count, input/output token geometry and the published TP/DEP topology can be represented explicitly~\citep{nvidia_trtllm126post3}. End-to-end output throughput, including prefill and decode, is the validation observable. At provisional $\eta_{\rm HBM}=0.40$, six H200 cases are selected once using two mechanistic criteria: prefill fraction $\leq0.05$ and decode memory fraction $\geq0.95$. A single application-level HBM efficiency is then solved numerically through root-finding such that the mean log measured/model throughput residual over these six calibration cases is zero. The exact root was found to be 0.430196 and so for this work we set the production value $\eta_{\rm HBM}=0.43$.

\paragraph{Steady-state batching and utilisation}
LLM inference servers can improve efficiency by amortization - processing multiple user requests together, allowing the same model weights to be used for several token-generation operations as opposed to processing each request separately. Larger batches generally improve total token throughput, but require additional GPU memory for each request's key-value (KV) cache and can increase response latency~\citep{yu2022,kwon2023,agrawal2024}. SULLEE approximates this benefit of batching without simulating request arrivals, scheduling, or user-facing latency. Let $R_{\rm dec}(B)$ denote the aggregate decode throughput for a batch of $B$ requests, and let $B_{\rm cache}$ be the largest batch that fits within the available GPU cache memory. We define $B_{\rm amort}(q)$, the amortization-adjusted batch size  as:

\begin{equation}
B_{\rm amort}(q)
=
\min\left\{
B\leq B_{\rm cache}:
R_{\rm dec}(B)\geq qR_{\rm dec}(B_{\rm cache})
\right\}.
\end{equation}

The primary high-utilisation case uses $q=0.95$, selecting the smallest batch that achieves at least 95\% of the throughput at the maximum cache-feasible batch size. These are assumed operating conditions, not measured provider batch sizes or predictions of response latency.

\paragraph{Parallelism and communication}
Large LLMs may require multiple GPUs to accommodate their model weights and perform inference. SULLEE represents three established ways of dividing this work: tensor parallelism (TP) splits calculations within model layers, pipeline parallelism (PP) distributes successive layers across GPU groups, and expert parallelism (EP) distributes the specialist networks in mixture-of-experts (MoE) models \citep{rajbhandari2022,liu2025moe}. To reduce communication between separate compute nodes, the model keeps TP and EP groups within an eight-GPU node where possible, following established deployment principles \citep{shoeybi2019megatron,liu2025moe}. For $N$ GPUs, TP is the largest divisor of $N$ that also divides eight, with ${\rm PP}=N/{\rm TP}$ and ${\rm EP}={\rm TP}$ for MoE models (or unity for dense models). This is a representative deployment assumption, not a reconstruction of a commercial provider's system.

Distributing inference across GPUs introduces additional data transfers such as combining partial calculations (TP), sending tokens to and from selected experts (EP), and passing intermediate results between layer groups (PP). SULLEE estimates transfer times using 80\% of the nominal NVLink and inter-node network bandwidth to allow for non-ideal communication performance. Since some transfers can occur while GPUs perform other calculations, only 50\%, 50\%, and 75\% of the estimated TP, EP, and PP transfer times, respectively, are assumed to add to the active processing time. These bandwidth and overlap fractions are intended to be fair, representative assumptions rather than calibrated provider measurements. Further details are provided in Supplementary Equation~S6 and Table~S2D.

\subsection*{Sub-model 1 external serving benchmarks and industry-reference comparisons}
\label{sec:methods-validation}

Recent industry disclosures provide useful external comparison points, including OpenAI (ChatGPT) at 0.34 Wh per stated average query~\citep{altman2025}, Microsoft Research at 0.31 Wh for a median prompt~\citep{oviedo2025}, and Google (Gemini) at 0.24 Wh for a median text prompt~\citep{elsworth2025} across multiple models. Independent empirical measurements across diverse model architectures and GPU configurations support the existence of broad variation in inference energy~\cite{caravaca2025}. These provider disclosures are used as benchmarking not calibration targets for SULLEE's emergent median prompt energy requirement, because their workload geometries, statistical definitions, and serving conditions are not exactly known to the individual SULLEE requests.

Further benchmarking of the SULLEE model is also assessed using hardware specific total serving throughput benchmark measurements from the remaining unseen 13 H200 cases from the measured data set~\citep{nvidia_trtllm126post3} used in HBM calibrating $\eta_{\rm HBM}$. These throughput measurements are utilised as a held out test of the fit and to provide an uncertainty estimate. For held-out case $i$ we define a fit residual
\begin{equation}
r_i=\ln\!\left(\frac{R_{{\rm measured},i}}{R_{{\rm SULLEE},i}}\right), 
\end{equation}
where $R_{{\rm measured},i}$ is the externally measured throughput data point and $R_{{\rm SULLEE},i}$ is a total throughput estimated using the SULLEE model. 

The primary residual width is the RMS about zero over the 13 held-out H200 residuals, $\sigma_r=0.463198$. It is not converted into a post-hoc central energy multiplier. Instead, the institutional scale-up model draws a persistent mean-preserving log-normal model-form factor $Z$ with $\ln Z\sim\mathcal{N}(-\sigma_r^2/2,\sigma_r^2)$, so $\mathbb{E}[Z]=1$. One uncertainty factor is shared by all requests/streams within each hypothetical SULLEE cohort sampled in an institutional draw. Supplementary Equations~S7--S9 and Tables~S2D-F document the calibration, residual and fleet boundary; all 19 cases are reported in Table~S4.

\section*{Sub-model 2: Institutional inference task sampling}\label{sec:methods2}
We now describe how individual tasks are scaled up within SULLEE to estimate institution-level annual inference electricity requirement. Annual total energy and associated emissions are obtained by summing the modelled LLM-serving energy of all included human-initiated and machine-initiated workflow streams relevant to a given institution. The model coupling and relevant input and output streams are summarised in Figure~\ref{fig:figure1}.

The second sub-model translates inference request-level energy distributions from sub-model 1 into annual institutional outcomes by sampling plausible university populations, role-specific message intensity, active weeks, workflow-start rates, internal calls per workflow, tool-context additions, and non-chat AI activity. The purpose of this sub-model is not to predict the behaviour of any specific single university or institution with absolute numerical precision, but to compare plausible institution-scale adoption regimes on a transparent common basis.

While assigning specific values to the environmental footprint of AI inference at a given institution is difficult due to uncertain adoption, usage intensity, and workload composition, the resulting uncertainty can be propagated through the energy and emissions calculations using Monte Carlo simulation. Use of Monte Carlo simulation in this manner is an established approach in environmental and greenhouse-gas accounting for combining uncertain activity data and emission factors~\citep{ipcc2019uncertainties,ghgprotocol2011product}. We therefore sample plausible role-specific user activity, task mixtures, and AI-enabled workflows to estimate distributions of annual inference electricity and associated CO$_2$ emissions. These distributions are conditional on the assumed scenario inputs rather than empirical predictions of individual institutional outcomes.

We examine three scenarios representing different levels of institutional LLM adoption and integration: 1) the \emph{2026 observed-use} scenario represents largely self-directed adoption, consistent with recent Australian university research reporting widespread student GenAI use, much of it developed independently of institutional support~\citep{chung2026genai}l; 2) The \emph{2026 institution-supported} scenario represents more systematic adoption through institutional licences, training, and integration into teaching and administrative workflows, as demonstrated by existing university-wide AI deployments~\citep{asu2025deployment,csu2025ai}; 3) finally, the \emph{2030 high-integration} scenario explores a possible expansion towards more intensive and agent-driven workflows. Consistent with established scenario-based approaches to assessing uncertain future AI electricity demand~\citep{IEA_2025}, the \emph{2030 high-integration} scenario is intended as a forward-looking planning and stress-test case.

\subsection*{Institutional population and activity scenarios}
\label{sec:methods-population}

For each class of role $r$, such as students or staff, the annual number of user-originated LLM messages, $N_r$, in an institutional draw is
\begin{equation}
N_r
=
H_r\,A_r\,Q\,W_r,
\end{equation}
where $H_r$ is role headcount, $A_r$ is the weekly-active fraction of people in that role who use LLM tools, $Q$ is the institution-level mean number of messages per weekly-active user per week, and $W_r$ is the number of active weeks per year. 

Weekly-active fractions represent the proportion of students or staff who use an LLM service during a given week. These proportions vary with user role and institutional adoption, as demonstrated by recent higher-education surveys~\citep{gallup2026ai,qs2026genai}. To represent uncertainty in these fractions, we use beta distributions, which are bounded between zero and one and allow independent specification of the assumed mean and its dispersion~\citep{ferrari2004},
\begin{equation}
A_r\sim
{\rm Beta}(\mu_r\kappa_r,(1-\mu_r)\kappa_r),
\end{equation}
where $\mu_r$ is the assumed mean weekly-active fraction and $\kappa_r$ controls its uncertainty. These are scenario priors aimed at capturing the reasonable uncertainty range expected of university adoption rates. The weekly message intensity $Q$ and number of active weeks $W_r$ instead use triangular distributions to represent plausible lower, most likely, and upper values in the absence of sufficiently detailed institutional usage data~\citep{taylor1994}. A single $Q$ is drawn for each simulated institution because it represents uncertainty in institution-wide mean usage rather than variation between individual users.

The message-intensity ranges are informed by available measurements of ChatGPT use. Consumer-platform telemetry implies approximately 26 messages per weekly-active user in 2025~\citep{chatterji2025}, while university ChatGPT Edu reports document mean weekly usage of 34 messages at Virginia Tech and approximately 49 interactions per active user at Cedarville University~\citep{virginiatech2025pilot,cedarville2026ai}, for example. Guided by these observations, we adopt ${\rm Tri}(10,30,50)$ messages week$^{-1}$ for 2026 observed use and ${\rm Tri}(20,40,80)$ for 2026 institution-supported use. The 2030 high-integration case, ${\rm Tri}(30,80,130)$, deliberately represents a higher-intensity stress planning scenario. All three distributions bring uncertainty in the institution-level mean among weekly-active users, not individual message-count distributions. Role-specific request aggregation and task composition are defined in Supplementary Equations~S10--S11, with the underlying assumptions reported in Tables~S3A--S3B.

User task composition is represented by a six-component vector $\mathbf{w}$ per institutional Monte Carlo draw, describing the randomly sampled fractional breakdown of the six considered task types in this study: general chat, long document, code completion, grounded retrieval augmented generation (RAG), reasoning production, and deep problem solving. Rather than drawing six independent uniforms and renormalizing them, which would induce a different and poorly controlled joint prior, we sample uniformly from the feasible bounded simplex
\begin{equation}
\mathcal{W}
=
\left\{
\mathbf{w}:
\sum_{j=1}^{6} w_j=1,\;
\ell_j\leq w_j\leq u_j
\right\}.
\end{equation}
The adopted bounds for each of the six task types are chosen to prioritise user chat as the primary request task, with a range of 20\%-40\% of an institution's workload. Following chat, we designate long task requests (e.g. analyse a document) as the next most common task with a range of 15\%-30\% of workload requests. This is followed approximately equally by reasoning tasks~(12\%-25\%) and RAG~(10\%-25\%), and then coding~(5\%-10\%) and deep-reasoning~(2\%-10\%):
\begin{align*}
w_{\rm chat}&\in[0.20,0.40],&
w_{\rm RAG}&\in[0.10,0.25],\\
w_{\rm long}&\in[0.15,0.30],&
w_{\rm code}&\in[0.05,0.10],\\
w_{\rm reasoning}&\in[0.12,0.25],&
w_{\rm deep}&\in[0.02,0.10].
\end{align*}
All feasible compositions within this bounded simplex have equal joint density. The same user task-composition vector is applied to students, academic staff, and professional staff in a given draw because role- and discipline-specific task distributions are not presently measured at sufficient resolution.

\begin{figure}
    \centering
    \includegraphics[width=0.9\linewidth]{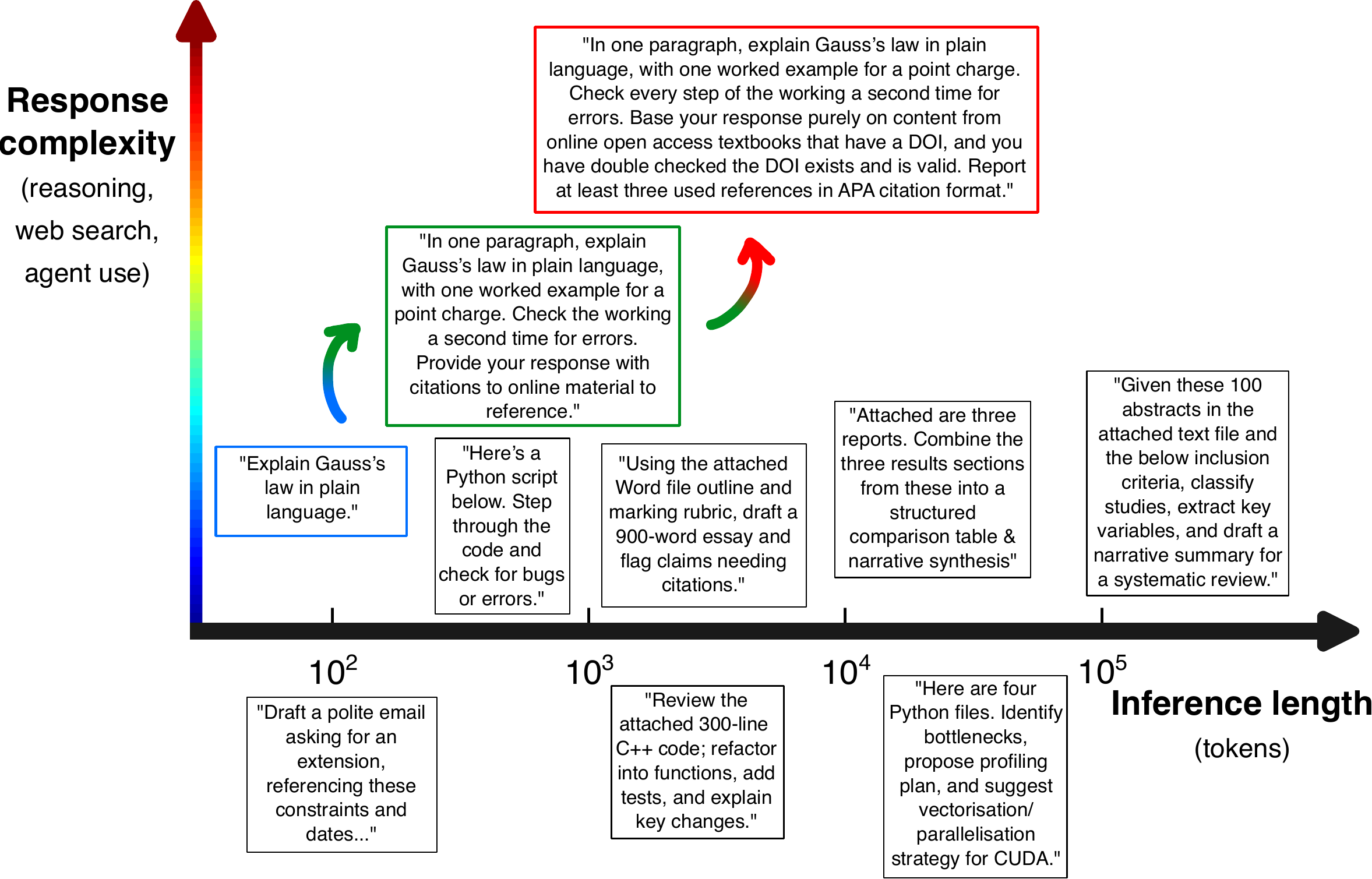}
    \caption{Representative prompt examples in the landscape of token use and prompt complexity}
    \label{fig:prompt_examples}
\end{figure}

\subsection*{Agentic and non-chat workflow model}
\label{sec:methods-agentic}

Rather than assigning one fixed energy value to all institutional requests, the scale-out model represents institutional frontier LLM tool use through heterogeneous task classes and separate conversational, agentic, and non-chat workflow streams. %

\paragraph{Workflow activity amplification}
AI agents can turn one user request into several LLM calls, for example when planning a task, searching for information, and interpreting results~\citep{han2026powercompression,liu2026agentic,bai2026agenttokens}. SULLEE represents these additional calls separately from ordinary human messages. For a collection of human-originated messages, $N_{\rm human}$, they are each counted once in the aggregate energy relevant model call load $N_{\rm workflow}$. However, some workflows are assumed to initiate agent activity, $N_{\rm agent\,calls}$, that themselves generate further LLM calls. For each simulated institution,

\begin{equation}
N_{\rm workflow}
=
N_{\rm human}f_{\rm workflow},
\end{equation}
and
\begin{equation}
N_{\rm agent\,calls}
=
N_{\rm workflow}C_{\rm workflow},
\end{equation}

where $f_{\rm workflow}$ is the number of model evaluation workflows started per human message and $C_{\rm workflow}$ is the number of additional agentic model calls per workflow. Thus, agent calls are added to the direct human-requested model evaluations. Workflow starts are not themselves counted again as separate calls. Both quantities use independent triangular scenario priors (Table~S3C), informed by evidence of multi-call agent activity but not fitted to university usage data~\citep{liu2026agentic,anthropic2025multiagent}.

Agents may also increase the amount of text processed when they retrieve documents, use tools, or pass results between steps~\citep{bai2026agenttokens,anthropic2025tooluse}. To represent this additional token input, a scenario-dependent fraction of agent calls receives extra input token context sampled as

\begin{equation}
\Delta T_{\rm tool}
\sim {\rm LogUniform}(3000,50000).
\end{equation}

This is a broad modelling range rather than a measured distribution. Its energy cost is calculated using SULLEE at additional context lengths of 0, 4k, 16k, 32k, and 50k tokens, instead of assuming a fixed electricity cost per token. Intermediate values are interpolated after removing small downward fluctuations in the estimated energy response; no extrapolation beyond 50k additional tokens is used.

This is intended to capture activity in additional agent calls, such as automatic code completions, embedded applications, and autonomous/API workflows. GitHub Copilot telemetry, for example, distinguishes user-submitted prompts from separate code-generation events~\citep{github2026metrics}. Embedded applications are represented as RAG LLM requests~\citep{ragpulse2025}. Electricity used directly by search engines, external tools, databases, and other non-LLM services is excluded; only their contribution to additional LLM calls and input context is captured. Numerical priors and further implementation details are reported in Supplementary Equations~S12--S13 and Table~S3C.

\subsection*{Student headcount and EFTSL conversion}
\label{sec:methods-eftsl}

Australian higher-education statistics distinguish student headcount from Equivalent Full-Time Student Load (EFTSL), which measures annual study activity rather than the number of  individuals~\citep{doe2025eftsl,doe2025staff}. Because our modelling is based on estimated AI use per person, institutional headcounts are used wherever available. When only student EFTSL is reported, we estimate headcount as a 

\begin{equation}
H_{\rm student}
=
\rho_{\rm EFTSL}\,N_{\rm EFTSL},
\qquad
\rho_{\rm EFTSL}\sim{\rm Tri}(1.25,1.38,1.50).
\end{equation}

This assumed range of the scaling factor $\rho_{\rm EFTSL}$ is informed by published university headcount-to-EFTSL ratios, from Australian university annual reports. The triangular distribution represents uncertainty about the conversion for an unspecified arbitrary sampled institution, rather than an exact fit distribution of Australian university enrolments. Similarly, when direct staff headcounts are unavailable, full-time equivalent (FTE) staffing is converted using

\begin{equation}
H_{\rm staff}
=
\rho_{\rm FTE}\,N_{\rm FTE},
\qquad
\rho_{\rm FTE}\sim{\rm Tri}(1.07,1.10,1.15).
\end{equation}

The assumed range of the scaling factor $\rho_{\rm FTE}$ is estimated based on typical published headcount-to-FTE ratios spanned by reported figures in university annual reports. A common conversion factor is applied to academic and professional staff within each simulated institution. This is a simplifying assumption, not a measured role-specific conversion. Because national casual-staff statistics report FTE but not headcount, the fallback is intended for non-casual staffing; casual-inclusive FTE requires separate treatment~\citep{doe2025staff}.

Where only aggregate staff numbers are available, we assume 42\% academic and 58\% professional staff. This follows the Australian Department of Education's 2024 current-duties FTE classification of employees in higher education institutions, which reports 51,390 academic and 71,199 non-academic FTE~\citep{doe2025staff}. These proportions are used as headcount approximations, and where available institution-specific headcounts take precedence in all cases. The direct-headcount priority and EFTSL/FTE fallback assumptions are reported in Table~S3A.

\subsection*{Electricity carbon-intensity and physical supply scenarios}
\label{sec:methods-carbon}

Equivalent CO$_2$ emissions are computed from the annualised energy use $\rm{E_{annual}}$ by simply scaling by a grid carbon intensity factor, $I_{\rm{grid}}$,
\begin{equation}
\rm{CO}_{2 \ {\rm{annual}}} = E_{\rm{annual}} \times I_{\rm{grid}}. 
\end{equation}

Using national or regional grid emission factors, $I_{\text{grid}}$, is well-justified and represents standard practice in carbon accounting \citep{ghgprotocol2015}. The GHG Protocol Scope 2 Guidance recommends location-based methods using regional grid factors when specific provider information is unavailable. The data centre location-based calculation converts annual electricity using published regional grid carbon intensities with $I_{\rm grid}$ expressed in kgCO$_2$e\,kWh$^{-1}$. For simplicity and aiding comparisons between scenarios of interest to this particular study, the published current grid factor is held constant across scenario years, including 2030. While some AI providers report renewable energy procurement, the actual electricity and emissions consequences depend on geographical locations of data centres, temporal matching of loads, or additionality of renewable energy generation or credits. Data centres require continuous power, and renewable energy certificates do not necessarily reduce marginal grid emissions~\citep{brander2018}.

The geographic distribution of data-centre infrastructure is highly uneven, with the United States hosting substantially more identified facilities than any other country, followed by major markets in Europe and Asia~\citep{stanford2026aiindex,statista2026datacenters}. For reference, Table~\ref{tab:emissions_intensities} details the average carbon intensities of countries or groups. Facility counts indicate geographic prevalence though not direct information of installed computing capacity, electricity consumption, or AI inference activity. Rather than specifying any specific data centre location we simply adopt four representative life-cycle greenhouse gas emission intensities, detailed also in Table~\ref{tab:emissions_intensities} as life-cycle generation-source sensitivities. They are used to examine the consequence of supplying the same modeled inference electricity from alternative generation systems. 

\begin{table}[htbp]
\centering
\caption{Electricity emissions life-cycle intensities for selected countries~\citep{ember2026co2}, and four electricity-generation scenarios~\citep{intergovernmental_panel_on_climate_change_ipcc_changing_2023}. The four electricity-source scenarios considered in this study are highlighted in italics for comparison with countries with highest data centre footprints.}
\label{tab:emissions_intensities}
\begin{tabular}{lr}
\toprule
\textbf{Country or generation scenario} &
\textbf{Emissions intensity (kg CO$_2$e\,kWh$^{-1}$)} \\
\midrule
\textit{Low-carbon mix    }& 0.025 \\
France         & 0.041 \\
Canada         & 0.192 \\
United Kingdom & 0.218 \\
\textit{50\% low-carbon / 50\% natural gas} & 0.281 \\
Germany        & 0.334 \\
United States  & 0.384 \\
Russia         & 0.447 \\
Japan          & 0.474 \\
Australia      & 0.518 \\
China          & 0.528 \\
\textit{Natural gas   }                    & 0.537 \\
India          & 0.671 \\
\textit{50\% coal / 50\% natural gas}       & 0.751 \\
\bottomrule
\end{tabular}
\end{table}

The life-cycle source sensitivities are fixed at 0.025, 0.281, 0.537, and 0.751~kgCO$_2$e\,kWh$^{-1}$ for, respectively, the low-carbon, 50\%-50\% low-carbon and natural-gas mix, natural-gas, and fossil-heavy 50\% coal/50\% gas cases~\cite{intergovernmental_panel_on_climate_change_ipcc_changing_2023}.  The 0.025~kgCO$_2$e\,kWh$^{-1}$ case is an intentionally rounded representation of a low-carbon mixed supply and should not be interpreted as a universal emission intensity for all forms of renewable electricity generation. The carbon conversion is defined in Supplementary Equation~S16; all operational and life-cycle factors, their boundaries, vintages, and derivations are reported in Table~S5.

\subsection*{Scope-3 contextual analysis}
\mbox{}\par
\label{sec:methods-scope3}

As universities integrate LLM tools into teaching, research, and administrative workflows, the associated emissions are most naturally treated as Scope 3 since they are derived from purchased services, where the GHG Protocol explicitly emphasises that credible Scope 3 reporting should use the best available data and be transparent about limitations~\footnote{\url{https://ghgprotocol.org}}. At the time of writing, in some deployment modes, suppliers are beginning to provide allocatable estimates: Google Cloud’s Carbon Footprint product~\footnote{\url{https://docs.cloud.google.com/carbon-footprint/docs/covered-services}} publishes covered-service lists that explicitly include Gemini API and Vertex AI, enabling account-level attribution for those workloads, but the output is structured around cloud usage reporting (e.g., by service and billing context) rather than institutionally meaningful units such as “per user” or “per query.”  In other common modes - especially SaaS-embedded assistants - this attribution problem is sharper: for example, at the time of writing, Microsoft’s own documentation for the Emissions Impact Dashboard for Microsoft 365~\footnote{\url{https://learn.microsoft.com/en-us/power-bi/connect-data/service-connect-emissions-impact-dashboard-microsoft-365}} states that Copilot for Microsoft 365 is not currently included, posing a challenge to direct vendor-supplied quantification of Copilot’s inference impact in day-to-day productivity use. 

Here, university Scope~3 disclosures were compiled manually from publicly available university annual reports, sustainability reports, greenhouse-gas inventories, and related institutional disclosures of 2024 or 2025 based emissions inventories. No common Australian higher-education dataset was identified in which Scope~3 emissions are reported using a harmonized reporting year, organizational boundary, category coverage, or calculation method. We obtained 21 useful data points and these observations are therefore treated as a manually curated contextual sample of publicly reported university Scope~3 inventories, rather than as a standardized or fully statistically representative sector dataset.

Institution-identifying labels are intentionally withheld because the purpose of our comparison is to provide broad context not rank or assess individual institutions. Reported totals are retained as published; missing categories are not reconstructed and differences in reporting boundaries are not retrospectively harmonized. Consequently, cross-university spread is wide and reflects both differences in underlying emissions and differences in reporting practice.

For university observation $u$, institutional usage scenario $s$, electricity-source case $g$, and institutional Monte Carlo draw $i$, the percentage of AI inference related Scope 3 emissions is
\begin{equation}
R_{usgi}
=
100\,
\frac{C^{\rm inference}_{usgi}}
{b_s\,C^{\rm reported\,Scope\,3}_{u}},
\end{equation}
where $b_s=1$ for both 2026 scenarios and $b_s=0.8$ for the 2030 high-integration sensitivity. The 0.8 multiplier is a simple target-informed accounting assumption, not a forecast. It simply is intended to compare a reasonable assumption of a growing inference load through technology adoption with an achievable 20\%-lower Scope~3 baseline, which many universities are transparently pursuing as part of broader net zero emissions targets. Published university targets demonstrate that reductions of this order or larger are plausible planning benchmarks, for example UNSW targets a 50\% reduction in total emissions by 2030~\cite{unsw_net_zero_2030}. 

For Figure~\ref{fig:figure6}, all 21 observations with paired EFTSL and reported baseline Scope~3 values are sampled by the full institutional Monte Carlo routine. The results combine modeled inference energy uncertainty with empirical heterogeneity of the non-harmonized Scope~3 denominators sampled from real institutions. So we reiterate that this is descriptive context rather than a probability distribution for the entire Australian university sector. Further details are discussed in Supplementary Equation~S17. The 21 retained observations with paired EFTSL and Scope~3 values are reported in Table~S6.

\subsection*{Uncertainty propagation and statistical analysis}
\label{sec:methods-statistics}

Monte Carlo methods are well-suited to this context, propagating uncertainty through models while providing distributional outputs instead of optimistic, false precision~\citep{ipcc2019uncertainties,ghgprotocol2011product, harrison2010}. We use Monte Carlo uncertainty propagation to sample and propagate uncertainty in model inputs to distributions of per-request and annual impacts.

We distinguish request-level aleatory variability from uncertainty in institution-level and model-level quantities. For task weights $w_j$ and conditional task energy moments $(\mu_j,\sigma_j^2)$, the mixture mean, $\mu_{\rm req}$, and variance ,$V_{\rm req}$, of an individual request are
\begin{align}
\mu_{\rm req}
&=
\sum_j w_j\mu_j,\\
V_{\rm req}
&=
\sum_j
w_j(\sigma_j^2+\mu_j^2)
-
\mu_{\rm req}^2.
\end{align}
For $N$ requests, expected aggregate energy is therefore
\begin{equation}
\mathbb{E}[E_N]=N\mu_{\rm req},
\end{equation}
whereas heterogeneous request-level aleatory variance scales as
\begin{equation}
{\rm Var}(E_N)=NV_{\rm req}.
\end{equation}
The simulation propagates this request-level term using the corresponding normal large-$N$ approximation.

This distinction is important because uncertainty in a task's arithmetic mean, inference hardware deployment family, institutional composition, or technology adoption level is persistent within an institutional draw and does not disappear as $1/\sqrt{N}$ simply because an institution generates many requests.

For institutional electricity accounting the arithmetic expectation of request energy is the relevant statistic because expectation is additive. The request-energy median is retained as a descriptor of a ``typical'' request but is not used as a substitute for the arithmetic mean when annual electricity is aggregated. Because arithmetic means of strongly right-skewed task distributions can be sensitive to finite Monte Carlo sampling of rare high-energy requests, such as tail-critical consumer/reasoning, research/reasoning, or research/deep-problem-solving type scenarios are independently re-evaluated using 20,000 conditioned SULLEE requests per set. 

The average energy required for each LLM task is estimated from a finite number of simulated requests, these averages have some statistical uncertainty. We estimate this uncertainty using bootstrap-derived ranges and, for each simulated institution, draw one adjustment factor for the average energy of each task and user cohort. The same factor is applied to all requests of that type within the institution. This represents uncertainty in the estimated average energy per task, which persists regardless of how many requests are made.

To identify which assumptions have the greatest influence on the estimated annual electricity use, we run one-at-a-time screening to repeat the simulations while varying one group of uncertain inputs at a time and holding the others fixed. For each group, we measure the resulting variation in the logarithm of annual electricity and compare it with the variation produced by the other groups of inputs. These comparisons provide a simple indication of which assumptions most affect the estimates, but do not measure exact contributions to total uncertainty because the combined interaction effects between inputs are not evaluated in this analysis.

Mixture moments, request-level aleatory propagation, persistent finite-SULLEE task-mean uncertainty are defined in Supplementary Equations~S14--S15 and reported in Tables~S3D--S3E. The separate systems-validation residual uncertainty is defined in Supplementary Equation~S8.

\subsection*{Software implementation, numerical validation, and reproducibility}
\label{sec:methods-reproducibility}

Model parameters and scenario envelopes are justified against published hardware specifications, industry disclosures or trends, and peer-reviewed literature, and the computational framework is released as open-source code to enable reproducibility and institutional adaptation. All calculations were implemented in Python.

Numerical Monte Carlo error was assessed separately from our three usage scenarios' uncertainty. The principal institutional calculations were repeated at increasing outer-loop sample counts, and convergence was evaluated for the arithmetic mean and P05, P50, and P95 of annual electricity in the representative 20,000-student cases. Production sample sizes were chosen only after these statistics changed by less than 1\% upon further increasing the number of draws. The production draw count, pseudorandom seed, convergence diagnostics, and input configuration are retained with the archived computational files.



\section*{RESULTS: SULLEE inference model benchmarking and validation}\label{sec:results}

As with any modelling exercise, it is essential to compare a model output against available reference data to ensure the output is reasonable and in line with external, independent information or data. Before applying the bottom-up inference energy model to an institutional scenario, we present benchmarking and validation results of the SULLEE model against available frontier LLM energy per query disclosures, as well as open source model measured inference throughput rates. 

\subsection*{Typical-query estimates describe the central workload but not aggregate expected energy}
\label{sec:results-distribution}
Available frontier model disclosures are scalar mean or median, or sometimes unspecific ``average'' or ``typical'' inference query energy usages provided by Google, Microsoft Research, and OpenAI. We note that in some cases there is ambiguity around what type of average is used, whether mean or median, whereas in other cases it is clearly stated as a median reported value. 

In this comparison we compare frontier model typical energy disclosures to the simulated closed-frontier reference populations sampled in the SULLEE model. We find SULLEE's individual-level request-energy distribution has a median of 0.297~Wh per request, and an arithmetic mean of 0.797~Wh per request (Figure 3B). The mean is 2.7 times the median, reflecting the strongly right-skewed workload energy distribution. Indeed, the P95 and P99 quantities for energy per task are 2.776~Wh and 8.624~Wh (Figure 3A,C), demonstrating why a typical-scalar request median statistic cannot be an additive estimator of aggregate electricity. Critically, recent industry values such as Google's 0.24~Wh median text prompt~\cite{elsworth2025}, Microsoft's 0.31~Wh median~\cite{oviedo2025}, and OpenAI's stated 0.34~Wh average ChatGPT query~\cite{altman2025}, all fall in the central region of this modelled distribution (Figure 3B). We stress that none of these industry values are used as an energy-calibration target in our modelling.  This agreement is a reassuring plausibility check of SULLEE as the providers' workload mixtures, statistical summaries, hardware, and accounting boundaries have not been used to define our model parameters. 

The secondary level of benchmarking is against measured throughput for inference, i.e. how many tokens per second can be processed through the hardware performing inference computation, for tested open source LLMs on a fixed hardware instance. In this case, as more data is available and open source LLM architectures and geometries are better known, we can perform a more rigorous comparison than the frontier LLM energy comparison.  While not a direct energy quantity, this throughput is nonetheless a valuable benchmark quantity as at fixed node power, energy per token is inversely proportional to throughput, so a shorter service time consumes less energy, as outlined in Methods and Supplementary Material. 

\begin{figure}
    \centering
    \includegraphics[width=0.99\linewidth]{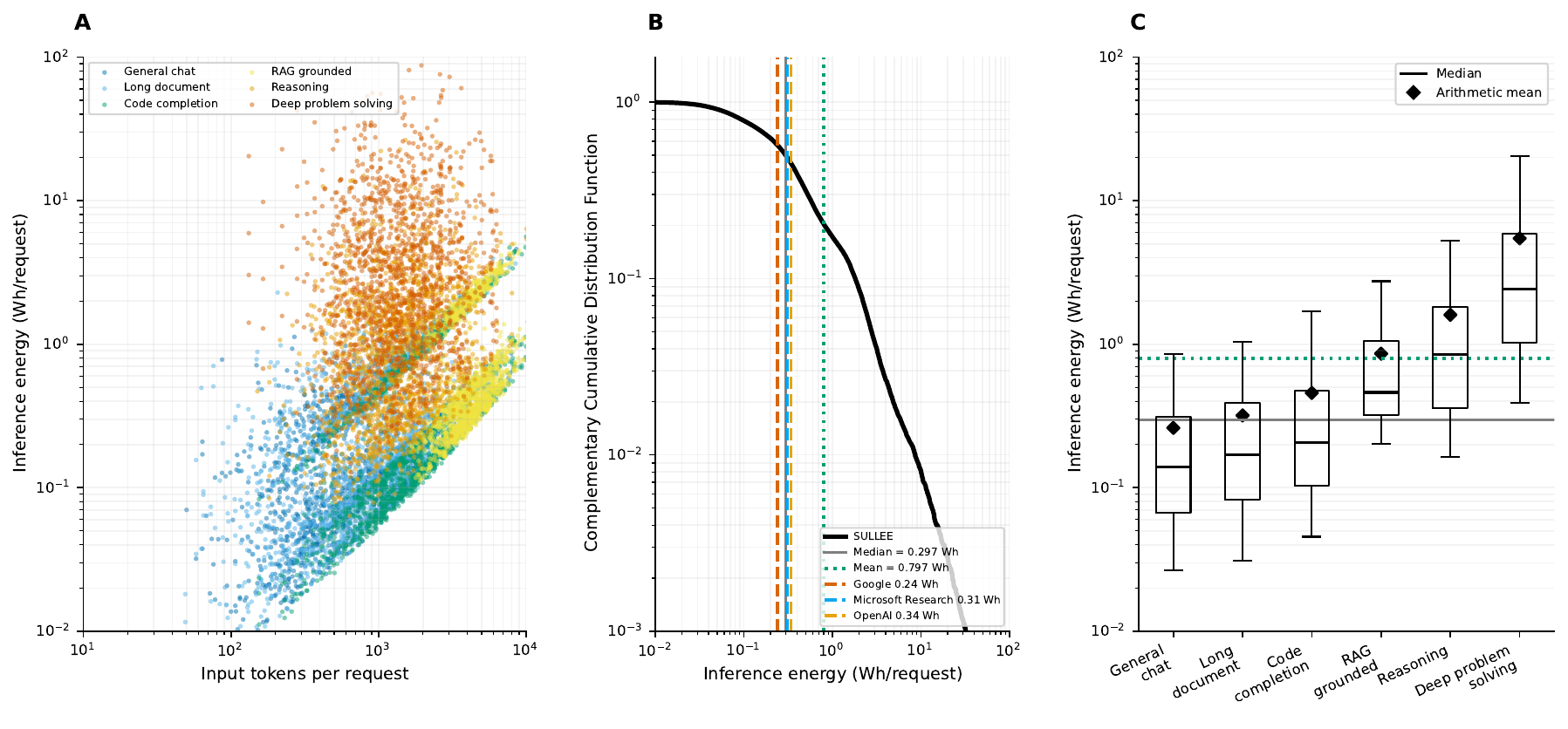}
    \caption{\textbf{A.} Inference request energy scatter plot of six representative task types under the frontier model scenarios sampled over 50,000 example queries. \textbf{B.} Complementary cumulative distribution function, or survival function, of task inference energy with vertical lines denoting mean~($\cdot \cdot \cdot$) and median~(--). Dashed vertical lines~(- - -) denote external industry reference values. \textbf{C.} Inference energy statistics for the six representative task types under the frontier model scenarios. Boxes span the 25th-75th percentiles, horizontal lines indicate medians, whiskers extend to the 5th-95th percentiles, and diamond markers denote arithmetic means. Horizontal lines denote mean~($\cdot \cdot \cdot$) and median~(--) values of~Figure~\ref{fig:figure2}B.} 
    \label{fig:figure2}
\end{figure}

\subsection*{External memory throughput benchmarks support model validity}
\label{sec:results-validation}

We use 19 public measurements of LLM inference throughput performance for Hopper era benchmark cases spanning GPT OSS 120B, Llama 4 Maverick 400B, Qwen 3 235B, and DeepSeek R1 671B. Six cases selected a priori as decode and memory dominated were used to baseline a representative HBM roofline efficiency, which we round and fix at $\eta_{\rm HBM}=0.43$, as discussed in the Methods section. In order to provide a benchmark of the overall SULLEE approach and assurance of the suitability of this model parameter, the remaining 13 H200 cases are held out from that calibration fit and are tested unseen. Throughput rates of 12 of those 13 (92.3\%) are within a factor of two of measured end-to-end output throughput, across a roughly 65-fold system throughput range, while the geometric measured/model ratio is 1.042.

The RMS log residual error of these 13 held out cases, $\sigma_{\ln}=0.4632$, is propagated when performing the follow-up institutional scale-up model as a mean-preserving log-normal model-form uncertainty that is persistent within each simulated SULLEE cohort. We note that in our modelling the input token context lengths are treated conservatively based on reasonable operating limits of contemporary open source LLMs measured with an upper ceiling of approximately ~10$^5$~tokens. For context lengths beyond those benchmarked to the public data of these open source model comparisons, especially $\geq128$k tokens, any prediction is extrapolative and are not considered as empirically validated.

\begin{figure}
    \centering
    \includegraphics[width=0.9\linewidth]{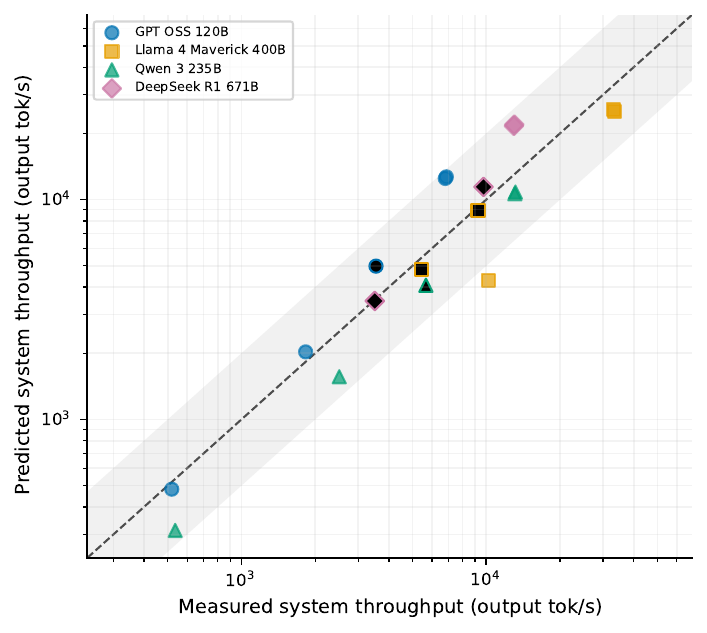}
    \caption{Goodness of fit of theoretical compute throughput versus recorded measurements of multiple open architecture models used as both an upstream high bandwidth memory efficiency calibration, and a held-out serving validation benchmark for the SULLEE model overall. Black-filled markers denote calibration set, colourised markers are held-out validation set. For reference, a shaded band denotes a factor of two about the 1:1 line.}
    \label{fig:figure3}
\end{figure}

Importantly, this exercise validates the throughput/runtime component of SULLEE under well defined public model, hardware, token-geometry, and parallelism cases. We then use the held-out throughput residuals to quantify reasonable inference model-hardware variations and uncertainty in the institutional model.

\subsection*{Task heterogeneity and reasoning workloads generate the long request-energy tail}
\label{sec:results-task-tail}

The range of variable inference task types, and token requirements observed and incorporated in SULLEE, naturally lead to the observation that energy required for individual-level prompts is strongly heterogeneous (Table~\ref{tab:task_energy}). Mean energy rises from 0.261~Wh for general chat to 1.601~Wh for reasoning-production workloads and 5.464~Wh for deep problem solving; the mean deep-problem-solving request is therefore approximately 20 times the mean general-chat request (Figure 3C). For the range of samples demonstrated in Figure 3, sorting query energies across the entire population of sampled hyperscaler class frontier LLM architectures, we observed that the highest-energy 1\%, 5\%, and 10\% of requests contributed 22.6\%, 43.8\%, and 57.8\% of total modelled request electricity, respectively.  These trends are not new and certainly not unique to this present study, as prior analysis has established this disproportionate impact of relatively rare but intensive inference tasks on energy budgets~\cite{oviedo2025}.  

However, by presenting the task breakdown and statistics in Table~\ref{tab:task_energy} we demonstrate the SULLEE's alignment with this established consensus, and also to provide approximate quantitative comparisons of the smallest, most frequent tasks to some long tail resource-intensive tasks. 

\begin{table}[htbp]
    \centering
    \caption{Statistical summary of energy consumption by inference task type in the representative frontier model request population modelled via SULLEE. The mean represents the average energy used in aggregate electricity calculations, the median describes a typical request, and P95 indicates the energy below which 95\% of modelled requests in each task category fall.}
    \label{tab:task_energy}
    \begin{tabular}{lccc}
        \toprule
        \textbf{Task} & \textbf{Mean (Wh)} & \textbf{Median (Wh)} & \textbf{P95 (Wh)}  \\
        \midrule
        General chat         & 0.261 & 0.140 & 0.856  \\
        Long document        & 0.319 & 0.169 & 1.042  \\
        Code completion      & 0.458 & 0.208 & 1.700  \\
        RAG grounded         & 0.859 & 0.462 & 2.748  \\
        Reasoning            & 1.601 & 0.852 & 5.297  \\
        Deep problem solving & 5.464 & 2.424 & 20.448  \\
        \midrule
        \textbf{All tasks}            & 0.797 & 0.297 & 2.776  \\
        \midrule
        \textbf{Open AI average\cite{altman2025}}            &   \hspace{35mm}0.34* &  &  -  \\
        \midrule
        \textbf{Google\cite{elsworth2025}}            &  - & 0.24 &   -   \\
        \midrule
        \textbf{Microsoft Research\cite{oviedo2025}}            & -  & 0.31 &  -   \\
        \bottomrule
    \end{tabular}
\end{table}

\section*{Institutional case study}\label{sec:results-case}
Here we present the results of scaling-up the generalised inference energy modelling of SULLEE towards a case study of inference workloads at Australian university institutions.

\subsection*{Institution-level AI energy demand is not driven by the typical  request}
\label{sec:results-framework}

To begin this section, we re-iterate a key concept of inference energy accounting. A majority of institution-level AI inference requests will be computationally cheap, such as the median chat request energy of~0.14~Wh in Table~\ref{tab:task_energy}. The largest energy draws originates from a minority of computationally intensive requests, such as reasoning or deep problem solving with medians of~0.85~Wh and~2.4~Wh respectively.  Indeed, the mean deep-problem-solving request requires approximately a factor of 20 times the energy of the most common brief general chat query. Consequently as the technology is adopted and ingrained across workplaces, growth in reasoning, long-context, corporate document input, and multi-stage workloads can increase aggregate energy requirements even without a proportionate increase in the number of visible user interactions. This observation is qualitatively consistent with recent independent production-oriented analysis which demonstrated that even a modest fraction of long reasoning queries can materially increase frontier-class LLM inference energy requirements~\cite{oviedo2025}.

\subsection*{Visible user prompts will not represent the full number or distribution of inference requests}
\label{sec:results-institutional}

As institutional adoption of LLM tools grows, the number of distinct user-submitted prompts and underlying agent or tool LLM calls are expected to diverge as workflow integration of LLM chat or other services increases~\cite{ragpulse2025,mcdonald_apostles_2024,stanford2026aiindex}. 

\begin{figure}
    \centering
    \includegraphics[width=0.9\linewidth]{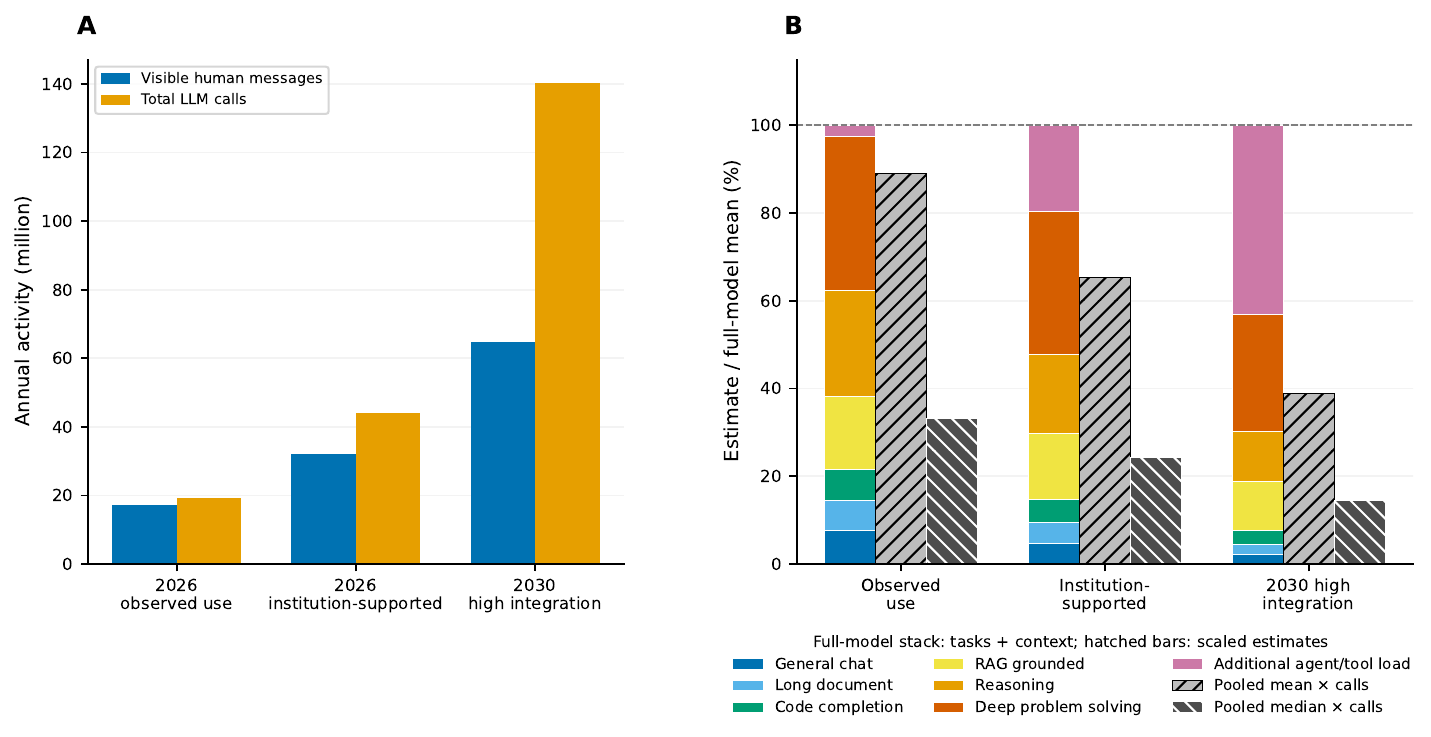}
    \caption{\textbf{A.} Workflow activity amplification - Annual user-originated messages compared with the total number of modelled LLM inference calls across all six activity streams. The total includes the human-message baseline and additional agent or tool calls. \textbf{B.} The shortfall introduced by scaling annual inference query demand by a single simplified per-query energy median statistic or using the mean energy per-query, compared to full heterogeneous accounting. Full breakdown includes stacked representation of task contribution for comparison.}
    \label{fig:figure4}
\end{figure}

Presented in Figure~\ref{fig:figure4}A, our modelling estimated mean annual \textit{user-originated} messages of 17.17, 32.05 and 64.63 million in the 2026 observed-use, 2026 institution-supported, and 2030 high-integration scenarios, while the corresponding \textit{total} modelled LLM calls are 19.38, 44.15 and 140.15 million. So, in the modelled high-integration 2030 case the model-call count is more than twice the visible message count initiated by users.

Expressed as a model-call amplification ratio, these totals correspond to approximately 1.13, 1.38, and 2.17 per visible human message, respectively. The increase is therefore not simply a proportional expansion of conventional chat activity: increasing workflow integration changes both the number and composition of inference calls. Multi-step agentic systems provide an independent systems precedent for distinguishing visible user requests from the underlying sequence of model calls ~\cite{han2026powercompression}. This further underscores the importance of heterogeneous task accounting.

Decomposing the representative 20,000 student institution scenario by service stream shows a qualitative transition as AI integration increases. Conventional conversational use accounts for approximately 87\% of modelled electricity in the 2026 observed-use case, but only approximately 20\% in the high-integration 2030 scenario. Conversely, agent workflows increase from approximately 8\% to approximately 78\% of modelled electricity. In the high-integration 2030 calculation, added tool and workflow context alone contributes approximately 43\% of annual inference electricity, whereas ordinary general-chat activity contributes only approximately 2\%. Thus, the principal scaling risk represented by this scenario is not simply assuming simply more chat requests, but a change in the computational structure of AI use toward multi-call, context-rich workflows as a non-trivial characteristic that must be actively considered in energy accounting.

\subsection*{Institutional aggregation amplifies errors from single-number per-query accounting}
\label{sec:results-aggregation}

The modelled annual inference energies, with uncertainty, attributed to 20,000 student, 40,000 student, and Australian total university groupings are presented in Figure~\ref{fig:figure5}A.For the representative 20,000 student university proxy, mean annual community-boundary electricity is 17.3~MWh (P05--P95: 3.1--47.1~MWh) for 2026 observed use, 53.9~MWh (9.8--144.8~MWh) for 2026 institution-supported use, and 287.4~MWh (41.5--830.7~MWh) for 2030 high integration. For comparison the group of larger universities modelled as the 40,000 student scenario proxy means are 34.9, 108.4, and 575.9~MWh, an approximate doubling from the 20,000 student case. Scaling the scenarios of LLM inference to the aggregate indicative Australian university sector population gives mean annual inference electricity of 1.46~GWh (P05--P95: 0.26–3.98~GWh) for 2026 observed use, 4.51~GWh (0.83–11.89~GWh) for 2026 institution-supported use and 24.08~GWh (3.43–69.12~GWh) for 2030 high integration.

The uncertainty depicted by these institutional P05--P95 ranges should not be interpreted as request-sampling noise or bad statistics accumulated over millions of independent queries. Request-level aleatory uncertainty averages down with large request counts, whereas the uncertainty simply from the high-dimensional number of possible plausible model configurations with architecture family, system-model residuals, task composition, or adoption rate is persistent across many requests within an institution. 

A key take away result of this work is highlighted in Figure~\ref{fig:figure4}B that simply multiplying total call count by a pooled typical request-energy median captures only approximately 33\%, 24\% and 14.5\% of the full modeled annual energy across these three institutional scenarios. Even multiplying by the pooled arithmetic mean captures approximately 89\%, 65\% and 39\%, because workflow streams can have different task mixes and context increments. The result is therefore not merely a mean-versus-median issue: accurate accounting also requires the composition of the calls being generated.

\subsection*{Provider electricity supply determines the carbon consequence of inference demand}
\label{sec:results-energy-source}

For a given scenario of institution-level electricity demand, the carbon-intensity of the electricity supply powering data centres serving inference is the major driver of attributed institution-level carbon emissions. As outlined in the Methods, in this study we employ four scenarios based on established emissions accounting modelling power generation methods~\cite{intergovernmental_panel_on_climate_change_ipcc_changing_2023}: a low-carbon (0.025~kgCO$_2$e\,kWh$^{-1}$) scenario, a 50\% low-carbon/50\% natural-gas (0.281~kgCO$_2$e\,kWh$^{-1}$) scenario, a natural-gas (0.537~kgCO$_2$e\,kWh$^{-1}$) scenario, and a fossil-heavy 50\% coal/50\% gas (0.751~kgCO$_2$e\,kWh$^{-1}$) scenario. 

\begin{figure}
    \centering
    \includegraphics[width=0.9\linewidth]{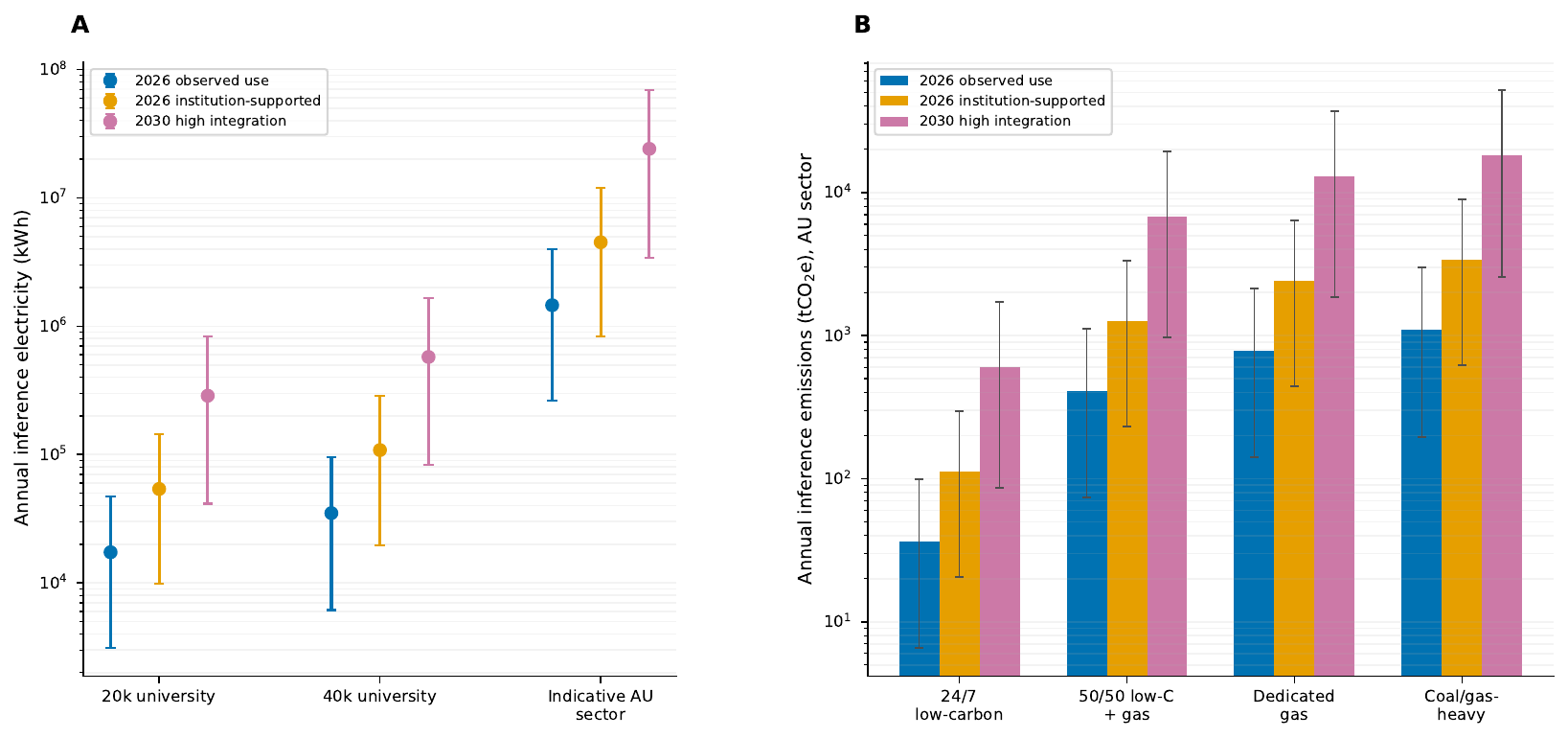}
    \caption{\textbf{A.} Institutional proxy and total indicative sector electricity demand associated with LLM inference activity under the conservative 2026, high-uptake 2026, and projected high-integration 2030 scenarios. \textbf{B.} Modelled carbon emissions attributed to the total Australian university sector dependent on carbon emissions intensity of data centre power generation method.}
    \label{fig:figure5}
\end{figure}

These four emissions sensitivities are applied to the conservative 2026, high-uptake 2026, and projected high-integration 2030 scenarios in Figure~\ref{fig:figure5}(B). 
For each data series presented in Figure~\ref{fig:figure5}, the plotted P05--P95 band propagates uncertainty in modelled inference demand only, for simplicity an uncertainty in the source-specific carbon emissions factor is not included.

Recent evidence indicates that energy use for a fixed AI task has indeed fallen rapidly as hardware and software have improved. However at the same time, AI uptake, reasoning depth, and agentic use have expanded. The IEA therefore identifies efficiency, adoption, and changing use-case intensity as separate determinants of aggregate AI electricity demand, and reports that AI-focused data-centre electricity increased even during a period of exceptionally rapid per-task efficiency improvement \citep{IEA_2025}. Rapid improvements in inference hardware or model efficiency do not necessarily imply a corresponding decline in aggregate electricity demand. This distinction is important for interpreting the 2030 high-integration scenario.

We can quantitatively unpack this to separate growth in workload demand from future improvements in serving efficiency. Let $g_{\rm eff}$ denote a uniform multiplicative improvement in end-to-end inference energy efficiency relative to the present day reference technology. For a fixed inference workload, the 20,000 student 2030 high-integration estimate is then
$$ E_{2030}(g_{\rm eff}) =
    \frac{287.4}{g_{\rm eff}}\ {\rm MWh\,yr^{-1}}. $$

If we then want to keep energy use unchanged relative to the 2026 institution-supported scenario mean of $53.9\ {\rm MWh\,yr^{-1}}$, the break-even efficiency improvement is $g_{\rm eff}=287.4/53.9=5.3$. While relative to the 2026 17.3~MWh conservative adoption scenario, a 287.4~MWh high-integration scenario in 2030 requires a near 17-fold increase in inference energy efficiency. A scenario considered recently by Oviedo et al.~\cite{oviedo2025} was potential for $8\times$ to $20\times$ efficiency improvements being achieved on a similar timeline, which are similar to the $g_{\rm eff}$ values required in our estimates.

The interaction between technical software or hardware efficiency gains and demand growth is especially important. Even an $8\times$ efficiency gain in energy use per equivalent workload, at the lower end of the combined efficiency opportunity estimated by Oviedo et al.~\cite{oviedo2025}, would leave the 2030 high-integration case at approximately 35.9~MWh~yr$^{-1}$ -- about twice the present observed-use scenario. Of course, a sufficiently larger efficiency gain can reverse this increase, but ultimately both efficiency improvement and workload growth must therefore be considered jointly. Rather than simply assuming that falling energy per query necessarily must imply falling aggregate electricity demand in future, we must be aware of the non-trivial and non-linear relationships presenting in generative AI uptake globally~\cite{CICCGlobalInstitute2026,Yuan2026,Henni2026}.

\subsection*{Inference emissions in the context of reported university Scope 3}
\mbox{}\par
\label{sec:results-scope3}

\begin{figure}
    \centering
    \includegraphics[width=0.9\linewidth]{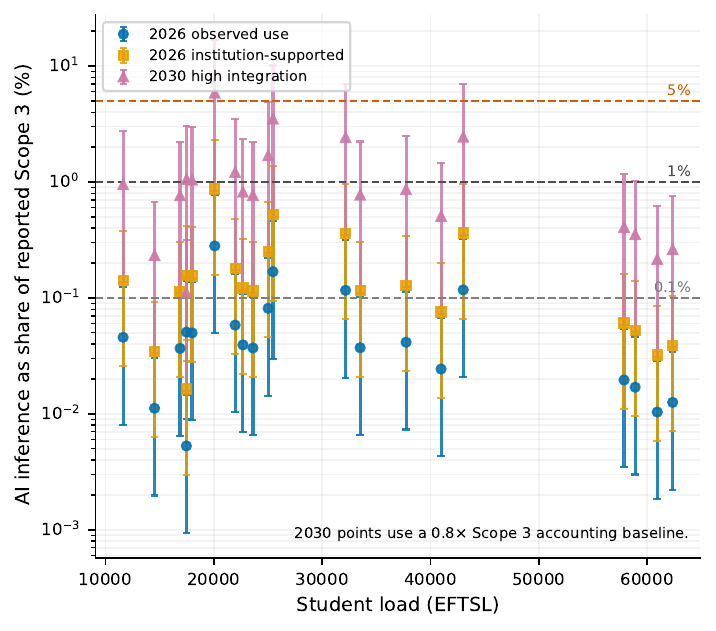}
    \caption{Modelled inference emissions as a fraction of reported Australian university Scope~3 inventories. The 2030 high-integration case uses a $0.8\times$ Scope~3 accounting baseline to model a reasonable, achievable reduction target broadly being pursued by the university sector; it is not a forecast of future emissions at any individual institution. Horizontal dashed lines denote 0.1\%, 1\%, and 5\% threshold values for interpretation.}
    \label{fig:figure6}
\end{figure}

Presented in Figure~\ref{fig:figure6} is Scope 3 data for 21 Australian university institutions of various sizes from 10,000 EFTSL to approximately 60,000 EFTSL. We estimate a range of percentage increases in Scope 3 emissions reasonably attributed to frontier LLM inference based on our modelling. Note that modelled annual AI inference emissions as a percentage of reported university Scope 3 emissions assumed a fixed Sydney regional electricity-grid intensity of 0.468~kgCO$_2$e,kWh$^{-1}$\footnote{Google Cloud Region Carbon data (2025) \url{https://github.com/GoogleCloudPlatform/region-carbon-info/blob/main/data/yearly/2025.csv}} across all three institutional usage scenarios. This was chosen to allow fair comparison between institutions, assuming they all use a hypothetical frontier generative AI service with data centres based in Australia.

Across the 21 university EFTSL/Scope~3 observations gathered, under assumed present 2026 workloads inference remains a small contextual fraction in the 2026 scenarios, typically on the order of 0.1\% of current Scope 3 values. Under the 2026 high integration scenario however, within uncertainty, some institutions demonstrably achieve non-trivial increases to around 1\% of current reported Scope 3 emissions. 

While modest in the overall landscape of emissions, institutions need to be aware of how small, often unseen, changes in frontier LLM usage can shift attributed emissions from a negligible rounding error to a distinct and considerable fraction of total Scope 3 emissions accounting.

Looking into the future of LLM technology and adoption is very difficult given the rate of change of the technology, it is nonetheless worthwhile to consider a future 2030 scenario based on the trajectory of uptake and technology growth. In this estimated 2030 uptake scenario, the median fraction of Scope 3 growth attributed to frontier LLM inference is approximately 1\% of baseline 2030 estimated Scope 3 emissions. Indeed, for 8 of 21 universities with data available the proportion of Scope 3 emissions was above 1\% in the 2030 scenario, and within reasonable uncertainty some comfortably approach 5\%.


\section*{DISCUSSION}
\mbox{}\par
\label{sec:discussion}
\label{sec:discussion-functional-unit}

A key outcome of this study is to demonstrate SULLEE provides a credible basis for both generalised frontier LLM inference task energy modelling and also institution inference energy budget estimation by natively factoring in inherent variation and uncertainty in model configurations, inference workloads, and usage rates rather than a single undifferentiated ``energy per query`` scalar number.

The principal strength of SULLEE is that it links physically meaningful features of frontier level LLM transformer inference (e.g. phase-aware prefill/decode modelling, sparse activation, grouped-query attention, roofline-bounded throughput, explicit data-centre overhead) to institution-relevant decision making variables in a transparent way with uncertainty propagation. 

We demonstrated the applicability of the modeling under uncertainty, where the SULLEE model's central median request-energy (0.297 Wh) was comparable with recent disclosures from OpenAI (0.34 Wh), Microsoft Research (0.31 Wh), and Google (0.24 Wh). An observed wide and strongly right-skewed distribution of energy-per-inference task reflects real variation of model architectures and optimisations, hosting configurations, workload distributions, statistical definitions, and system boundaries used by different frontier model operators. Systematic empirical measurements confirm that inference energy varies substantially across model architectures, quantisation levels, and hardware configurations even for comparable prompt lengths~\cite{caravaca2025}. This further reinforces our recommendation for distributional and workload-aware energy accounting instead of simply scaling up a single reported number for large scale institutional inference-energy accounting.

If we recall the scenario of university students asking 100 questions, posed by Dunn and Zhao~\cite{DunnZhao2026} in a recent Washington Post article, this example illustrates the distinction between a typical (median) interaction and the total energy attributable to a collection of interactions. For 100 model calls drawn from the SULLEE reference workload, multiplying the median by 100 gives 29.7 Wh, whereas the expected total by accounting over a distribution of query workloads, is 79.7 Wh because rare but higher-energy calls contribute disproportionately. Institutional adoption introduces a further change when a user submits a question, which may initiate several model calls or add document or web search retrieval to the compute tasks. This is reflected in Figure~\ref{fig:figure4}B, where multiplying all modelled calls by the reference mean of 0.797 Wh recovers about 89\% of annual electricity in the 2026 observed-use scenario, and only 39\% in the 2030 high-integration scenario evaluated at reference serving efficiency. For institutional procurement, the shows that useful disclosures, or ideally data telemetry, report information on the number of model calls generated by styles of user query, and their mean energy under representative use.

\subsection*{Standard LLM energy efficiency ratings are urgently needed}
\label{sec:discussion-ratings}
The substantial variation in inference energy across models, tasks, hardware configurations and deployment conditions highlights the need for greater transparency from frontier model providers. Independent initiatives such as the AI Energy Score~\cite{luccioni2025aienergyscore} and ML.ENERGY~\cite{mlenergy2025} benchmark have demonstrated that task-specific inference energy can be measured and communicated through standardised benchmarks, including readily interpretable efficiency ratings such as a one to five star rating. However, independent measurement remains considerably easier for openly available models than for proprietary frontier systems, for which the information required to reproduce or verify energy measurements is generally controlled by the provider. Frontier providers should therefore routinely disclose independently verifiable, task-specific inference-energy and efficiency metrics for their production services, enabling consumers and organisations to consider energy efficiency alongside capability, cost, or latency when making decisions in selecting AI systems. Such disclosure aligns with emerging regulatory direction, for example the EU AI Act requires general-purpose AI providers to document known or estimated model energy consumption~\cite{eu2024aiact}, while the European Commission is developing measurement methods that could underpin a future AI energy and emissions consumer-facing~\cite{ec2026aienergylabel}.

Such an efficiency metric should not, however, be interpreted in isolation. Comparisons are meaningful only when the functional unit is sufficiently controlled: at minimum, the task and evaluation set should be specified and energy reported alongside model quality, output geometry, latency or throughput, hardware, and the serving boundary. Otherwise, a less capable model, or one that systematically produces shorter outputs, can appear ``more efficient'' simply because it performs less useful computation. Existing benchmarking efforts therefore standardise tasks and report energy in conjunction with performance or service metrics \cite{mlenergy2025,luccioni2025aienergyscore}. For procurement, an energy rating should consequently be treated as one dimension of a multi-objective model-selection decision rather than as a standalone ranking of model quality.

At minimum, a provider-level disclosure should identify the inference functional unit; the input- and output-token geometry; whether the reported statistic is a mean, median, or other quantile; the model and hardware class; relevant serving or utilisation conditions; the included IT and facility boundary; PUE and any fleet- or reserve-capacity allocation; and the geographic and temporal basis of any associated carbon factor. Without these fields, numerically similar Wh~query$^{-1}$ values can represent materially different workloads and accounting boundaries. Full-stack production measurements~\cite{elsworth2025} and emerging European measurement work~\cite{ec2026aienergylabel} support movement toward this more explicit, system-boundary-aware form of disclosure.

\subsection*{Implications for organisational energy demand and Scope 3 reporting }
\label{sec:discussion-agentic}

Here, we found that the Scope~3 comparison does not imply that AI inference becomes a dominant component of Australian university emissions. Its relevance is that a single externally supplied digital-service class can move from effectively negligible in the 2026 scenario to a separately measurable 1-5\% scale procurement exposure for some institutions under high integration and uptake trajectories. As universities reduce other value-chain emission sources, failure to obtain supplier-specific AI energy and emissions data could therefore become increasingly consequential for inventory accuracy.

The strong dependence of institutional inference energy on usage intensity, task composition, and workflow amplification implies that institutional AI governance policies - including guidelines on appropriate use, awareness of query complexity, and deliberate selection of model tier (e.g.\ using lighter models for simple tasks) - constitute direct levers for emissions management. This represents an unusual case where behavioural and procurement policy can demonstrably influence Scope~3 emissions with relatively low implementation cost.

In the absence of vendor-supplied allocation data, SULLEE and similar bottom-up tools can support interim Scope~3 scenario analysis and procurement assessment, provided the accounting boundary and uncertainty are explicit. Our present comparisons should not be read as an audited Scope~3 inventory: the university observations are heterogeneous in year, organisational boundary and category coverage, and the 2030 20\% reduction is a simple target-informed sensitivity rather than simulated sector trajectory. 

The constructive policy-relevant result is focused: as other Scope~3 categories are reduced while AI use grows, inference can move from a negligible share to a material procurement category for some institutions, making supplier-specific energy and emissions disclosure increasingly valuable.

For an institution procuring an externally hosted AI service, supplier-specific activity or product-level emissions information would be preferable to a generic spend- or sector-based proxy where reliable primary data are available. GHG Protocol guidance for purchased goods and services explicitly prioritises supplier-specific product-level data when suitable data can be obtained \cite{GHGProtocolScope3Calculation2013}. Bottom-up inference modelling is therefore most appropriately viewed as an interim planning and screening method where supplier allocation data are unavailable, not as a replacement for supplier-specific inventory data.

\subsection*{Implications for data-centre electricity supply, carbon accounting, and energy policy}
\label{sec:discussion-policy}

Our projections carry several implications for institutional sustainability governance and procurement. First, institutions procuring AI services should request per-query or per-user energy allocation data from vendors as a standard element of service-level agreements and sustainability due diligence. The emergence of prompt-level energy disclosures from Google~\cite{elsworth2025} and account-level carbon reporting through Google Cloud's Carbon Footprint product demonstrates that such transparency is technically feasible, even if not yet universally offered.

A second implication is that ``the carbon footprint of AI'' is not a single model property. The strong dependence of inference-related carbon emissions on the physical electricity supply suggests that data-centre hosting location, and the source of electricity required to power the data-centre, should be considered as procurement criteria alongside cost, latency, and data sovereignty. Across the physical supply sensitivities considered here, the carbon intensity of otherwise identical inference electricity differs by approximately a factor of 30 between the 24/7 low-carbon and fossil-heavy cases. 

Institutions with flexibility in vendor selection may therefore achieve meaningful emissions reductions by preferring providers with demonstrably lower-carbon physical electricity supply or credible continuous low-carbon energy matching. Consequently, two institutions making comparable use of similarly efficient AI systems can have materially different value-chain emissions depending on where and how those systems are supplied.

\subsection*{Energy accounting is complementary to, not an argument against, productive AI use}

Our results should not be interpreted as suggesting that AI should not be used or that that  AI inference is presently a large fraction of a national electricity budget. Relative to the estimated current 2026 observed-use scenario, AI electricity demand could increase by more than an order of magnitude as adoption, computational intensity, and workflow amplification increase to 2030. This is consistent with broader data-centre and energy discourse identified by other research, in that while aggregate fractional usage may remain modest the load growth is rapid and upstream demand is concentrated in particular facilities or electricity systems or generation sites~\cite{IEA_2025,IEA2026KeyQuestions}. This matters for institutional inference accounting as an emerging purchased digital service whose energy and emissions are presently difficult for customers to predict,  observe, manage, or compare.

The present analysis does not estimate the economic or productivity benefits of AI adoption and therefore cannot determine the net social or economic value of any of the modelled workloads. Some studies demonstrate that generative AI can generate productivity gains for some forms of knowledge work, although effects vary by task and worker~\cite{NoyZhang2023,BrynjolfssonLiRaymond2025,DellAcqua2026}. Our results should therefore not be interpreted as an argument that AI deployment is environmentally unjustified because it consumes electricity.

Rather, productivity and resource use are simultaneous outcomes of the same technology. If AI creates sufficient value to justify widespread adoption, improving the amount of useful work delivered per unit of electricity and carbon becomes more important, not less. Transparent energy measurement enables institutions to compare models, route simple tasks to less energy-intensive systems, evaluate suppliers, and obtain productivity gains with lower resource requirements. Ultimately, the relevant optimisation problem is not minimum AI use, but greater useful output per unit of energy and emissions.


\subsection*{Limitations of the study}
\label{sec:limitations}

Several limitations, assumptions, and boundaries of our present modelling should guide interpretation. The present analysis is restricted to operational electricity attributable to representative frontier-class LLM inference and excludes model training, development, and the embodied emissions of computing hardware. These additional impacts can be substantial, with recent work quantifying equivalent emissions from model experimentation and development, as well as the manufacturing of dedicated AI accelerators and their supporting hardware~\citep{luccioni2023,morrison2025,Schneider2025,Falk2026}. Quantifying these effects for commercial LLM services would additionally require information about model development, hardware lifetimes, resource sharing, and the total number of requests served over a model's operational lifetime. They are therefore excluded from the present operational-inference boundary rather than represented by an assumed universal per-request overhead.

Important uncertainty remains in proprietary model architectures of transformer class models, provider-specific scheduling and serving strategies, and the extent to which public benchmark serving conditions can approximate closed frontier deployments. Formal matched validation remains sparse, particularly at long context lengths and large multi-node deployments, while institutional task mixtures and agentic workflow rates remain scenario assumptions guided by available data and behaviours of emerging studies than measured university-wide prevalence. These limitations do not nullify the usefulness of the framework, but they do define the level of interpretation that is justified: the outputs should be read as transparent scenario estimates for planning and comparison, not as substitutes for audited provider disclosures which would be the most accurate, and preferred, reporting data source.

The institutional model estimates annual inference serving electricity and emissions rather than time-resolved facility load or peak power. It does not model arrival correlations or simultaneous agent branches, and therefore should not be used to infer grid peak demands from annual energy. Total agent calls and peak concurrent calls are distinct quantities in agentic systems \citep{han2026powercompression}.

Finally, the agentic and non-chat accounting boundary captures LLM inference calls and additional LLM context but not the energy of external search, retrieval databases, tool execution, speech recognition, or other application-layer services. Institutional AI-system energy could therefore exceed the inference-only quantities reported here where such services are material.

\subsection*{Outlook}
\label{sec:outlook}

The most valuable next steps for improving this approach are additional matched empirical measurements. Simultaneous disclosure of request geometry, model/hardware configuration, serving throughput, allocated power, and end-to-end energy across short-context, long-context, reasoning, and agentic workloads would permit direct validation of both the runtime and energy-accounting layers of our modelling. Institution-level telemetry on weekly-active use, task composition, and workflow amplification would likewise allow the present scenario priors to be replaced by locally measured activity distributions. Because both per-task efficiency and workload composition are evolving rapidly~\cite{IEA2026KeyQuestions}, SULLEE should be treated as a versioned planning model whose evidence layer is periodically re-evaluated as new production measurements become available.

\section*{RESOURCE AVAILABILITY}
\mbox{}\par

\subsection*{Data and code availability}
\mbox{}\par

A publicly available code package for standalone SULLEE and institutional scale-up calculations, outputs, and figure generation will be made available online following peer-review process. 


\section*{ACKNOWLEDGMENTS}
\mbox{}\par

RFH was supported by an Australian Research Council (ARC) Discovery Early Career Researcher Award no. DE2501000013
%


\section*{Declaration of generative AI and AI-assisted technologies in the manuscript preparation process}
\mbox{}\par

During the preparation of this work, the author(s) used ChatGPT and Claude to assist in manuscript draft review, code debugging, and code refactoring for public archival purposes. After using this tool/service, the author(s) reviewed and edited the content as needed and take(s) full responsibility for the content of the published article.


\section*{SUPPLEMENTAL INFORMATION INDEX}
\mbox{}\par

A supplementary information document is provided to partner this manuscript. Please refer to that document for more details on technical model equations, parameter priors, validation auditing, carbon accounting, and contextual Scope~3 data.


\clearpage
\bibliography{references}

\end{document}